\documentclass[fleqn,usenatbib]{mnras}
\usepackage{newtxtext,newtxmath}
\usepackage{ulem}

\usepackage{blindtext, subfig}

\usepackage[T1]{fontenc}

\DeclareRobustCommand{\VAN}[3]{#2}
\let\VANthebibliography\thebibliography
\def\thebibliography{\DeclareRobustCommand{\VAN}[3]{##3}\VANthebibliography}

\usepackage{graphicx}	
\usepackage{amsmath}	

\usepackage{pdflscape}

\usepackage{rotating}

\usepackage{hyperref}

\newcommand{\ceo}{C$^{18}$O} 
\newcommand{\nvssrg}{NVSS 012258+614815}
\newcommand{\iras}{IRAS 01202+6133} 
\newcommand{\Hii}{H\,{\sc ii}}
\newcommand{\Hi}{H\,{\sc i}}
\newcommand{\kms}{km\,s$^{-1}$}

\title[\Hi\ shell in S187]{Fragmented atomic shell around S187 \Hii\ region and its interaction with molecular and ionized gas}

\author[Zemlyanukha et al, ]{Petr Zemlyanukha$^{1}$, 
Igor I. Zinchenko$^{1}$, 
Evgeny Dombek$^{1,2}$, 
Lev E. Pirogov$^{1}$,
Anastasiia Topchieva$^{3}$, 
\newauthor
Gilles Joncas$^{4}$, 
Lokesh K. Dewangan$^{5}$,  
Devendra K. Ojha$^{6}$ and
Swarna K. Ghosh $^{6}$
\\
$^{1}$Institute of Applied Physics of the Russian Academy of Sciences, 46 Ul'yanov~str., Nizhny Novgorod 603950, Russia\\
$^{2}$Lobachevsky State University of Nizhny Novgorod, 23 Gagarin Ave, Nizhny Novgorod 603950, Russia\\
$^{3}$Institute of Astronomy, Russian Academy of Sciences, Pyatnitskaya str., 48 , Moscow, Russia\\
$^{4}$Centre de Recherche en Astrophysique du Qu\'{e}bec, D\'{e}partement de Physique, de g\'{e}nie physique et d'optique, \\Universit\'{e} Laval, Qu\'{e}bec, QC, G1V OA6, Canada\\
$^{5}$Physical Research Laboratory, Navrangpura, Ahmedabad 380009, India\\
$^{6}$Department of Astronomy and Astrophysics, Tata Institute of Fundamental Research, Homi Bhabha Road, Mumbai 400 005, India\\
}

\date{Accepted XXX. Received YYY; in original form ZZZ}
\pubyear{2021}

\begin{document}
\label{firstpage}
\pagerange{\pageref{firstpage}--\pageref{lastpage}}
\maketitle
\begin{abstract}
    The environment {of} S187, {a} nearby \Hii\ region (1.4$\pm$0.3 kpc), is {analyzed}. A {surrounding} shell has been studied in the \Hi\ line{, molecular lines, and also in infrared and radio continua.} {W}e report the first evidence of a clumpy HI environment in {its photodissociation region}. A background radio galaxy {enables the} estimation of the properties of {cold} atomic gas. The  {estimated atomic mass fraction of the shell}  is {$\sim$260~M$_{\odot}$}, the median spin temperature is $\sim$50~K, the shell  size is $\sim$4 pc with {typical} wall width {around} 0.2 pc. The atomic shell consists of $\sim$100 fragments. The fragment sizes {correlate} with mass with a power-law index of 2.39{-2.50}. The S187 shell has a complex kinematical structure, including the expanding quasi spherical {layer}, molecular envelope, {an} atomic sub-bubble inside the shell and two dense cores {(S187~SE and S187~NE)} at different stages of evolution. The atomic sub-bubble inside the shell is young, contains a Class II young stellar object and OH maser in the centre and the associated {YSOs} in the walls of the bubble. {S187~SE and S187~NE} have similar masses ($\sim$1200~M$_\odot$ and $\sim$900~M$_\odot$, respectively). S187~SE is embedded into the atomic shell and has a number of {associated objects} including high mass protostars, outflows, maser sources and other indicators of ongoing star formation. {No YSOs} inside S187~NE were detected, but {indications of} compression and heating by the \Hii\ region {exist}.
\end{abstract}

\begin{keywords}
stars: formation – ISM: clouds – radio lines:ISM - ISM: HII regions - ISM: evolution - ISM: bubbles
\end{keywords}
\section{Introduction}
The expansion of an \Hii\ region is thought to be a factor that induces star formation in the interstellar medium especially as a trigger of high mass star formation. Such a trigger can arise from the overpressurization of gas that bounds the region \citep{Thompson} leading to compression and fragmentation. The dynamics of this process is usually studied with C{\sc i}/C{\sc ii}/O{\sc i} observations \citep{CI,CII, CIII}. The atomic hydrogen production in a photodissociation region (PDR) is quite natural and is predicted by the existing models \citep{KirsanovaDenseAtomic,KirsanovaYetAnotherHIIModelling,AnotherHI}. Still, there are few observations of atomic shells around \Hii\ regions in the neutral hydrogen line \citep{S170HI,S173HI, Joncas}. Such rarity is caused by the challenges of \Hi\ observations. The \Hi\ emitting zone near the ionized region is usually garbled by the Galactic atomic gas in the line of sight. An estimation of the properties of the atomic gas is difficult. The \Hi\ spin temperature is correlated with the optical depth leading to uncertainties in estimations \citep{Saha}. Absorption profiles are essential for these estimations. Sometimes one is helped when {a} strong background continuum source occur{s} behind the studied object \citep{Nguyen}. 

To study the atomic hydrogen gas around a \Hii\ region, high resolution \Hi\ data should be obtained and compared with kinematics, morphology and properties of other species.

\begin{figure}
	\includegraphics[width=\linewidth]{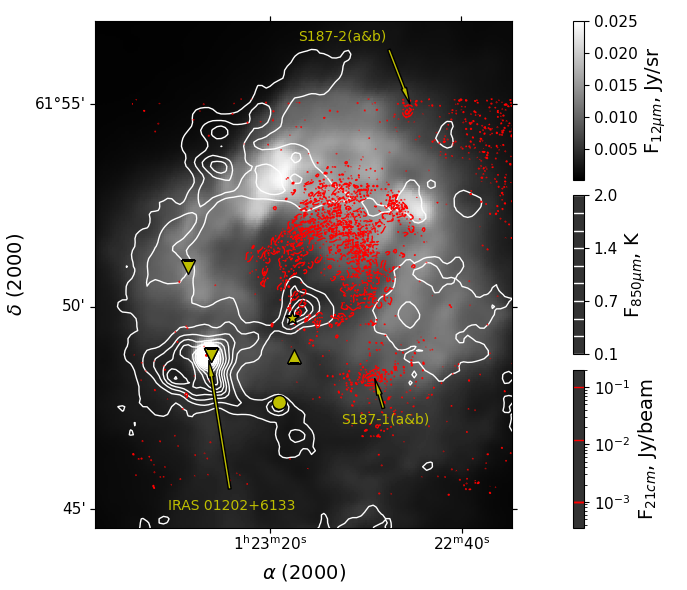}
	\caption{The structure of the S187  {complex}. The 1420 MHz GMRT continuum is in red {contours}, tracing the ionized gas. Bright red {spots} are related to radio galaxies (S187-1(a\&b) and S187-2(a\&b)){. The red contours near S187-1(a\&b) are caused by limited dynamical range. The noise on the top - right corner is caused by primary beam correction}
The 12$\mu$m WSSA image is {shown in} grayscale  tracing the dust in shocked gas. 
	White contours represent SCUBA~850$\mu m$ emission tracing cold dust in molecular material. 
	The water maser \citep{WM2007} is marked as a triangle and the {OH masers from \citet{Engels}} are marked as reversed triangles. {S187~NIRS~1 is shown as {a} circle {and S187 H$\alpha$ as the star}.}}
	\label{fig:continuum_spitzer}
\end{figure}

This paper presents such analysis for the S187 (Sh2-187, KR120, LBN 126.7-00.80) environment. This region is well-studied from the optical to the radio wavelengths. The S187 \Hii\ region (Fig. \ref{fig:continuum_spitzer}) is driven by a cluster of B0V stars with a photon flux of 10$^{47}$~{ionising} photons\,s$^{-1}$ \citep{Joncas} historically named as S187~B or BD52. The region is fairly young with an estimated age of $\simeq$5$\times$10$^5$~yr \citep{Joncas}. The distance estimate is 1.4$\pm$0.26~kpc from photometry \citep{photdist} and $\sim$0.9~kpc from Gaia~DR2 data and CO line analysis {of} associated large-scale structures \citep{altdist}. Since the uncertainty of S187's GAIA-based distance estimation is still not given we adopt the 1.4$\pm$0.26~kpc distance. The \Hii\ region is surrounded by a molecular and atomic gas shell \citep{Joncas, Arvidsson}. A number of young stellar objects were detected towards the shell \citep{Kang2017,Richards}. Three sources with maser emission were also detected including two OH \citep{Engels} masers and an H$_2$O maser \citep{Arcetri} associated with a CH$_3$OH maser \citep{Slysh}, indicating ongoing star formation in the shell. Two compact radio sources were reported as known radio-galaxies \citep{snell,Richards}. They were misinterpreted as UC~\Hii\ regions \citep{Israel} and have historical names S187-1(a\&b) (\nvssrg, ABG85 94) and S187-2(a\&b) (NVSS 012250+615451).

The formation of \iras\ ($M\sim$21~M$_\odot$, $L\sim$5600~L$_\odot$) is argued to have been triggered by the expansion of the S187 \Hii\ region \citep{Kang2012}. A pre-main-sequence object S187H$\alpha$ was discovered by \citet{Zavagno} with a mass of 25$\pm$12~M$_{\odot}$ {nearby to \iras.}.  A $^{12}$CO outflow is associated with S187H$\alpha$ \citep{Zavagno}. A  {molecular hydrogen} outflow {driven by T Tauri star S187 NIRS 1 (S187:SCP 2) reported in  \citet{Salas}.} {T}he H$_2$O \citep{Henkel86,Han95,Palagi,Arcetri} and {class II} CH$_3$OH maser \citep{Slysh} sources {are located} nearby{.}{The existence of such objects} is evidence of ongoing  star formation in the shell. An OH maser \citep{Engels} {was} also detected to the north of \iras, but we are not aware of any special study of the associated YSO. The dense molecular gas properties were discussed by \citet{Zinchenko2009, Kang2012} (see also references therein). The molecular part of the shell was observed in $^{12}$CO, \ceo\ and $^{13}$CO (1--0) and discussed in \cite{Arvidsson}, with the total mass estimate of about 7600~M$_{\odot}$.

A multifrequency study of S187 and its environment was performed by \citet{Joncas}, including \Hi\ data with 1$^\prime$ angular resolution and 1.32~\kms\ spectral resolution at 4.9~K sensitivity. According to \citet{Joncas}, the \Hii\ region is interacting with the surrounding neutral gas. The ionised region is surrounded by the shell containing the \Hi\ counterpart. {The a}tomic gas {distribution} is inhomogeneous, has a systematic velocity gradient and the total mass is 70~M$_{\odot}$. The associated YSO population is reported by \citet{Kang2017,snell}. Two types of dust populations with grains larger than 0.004 $\mu$m were detected in infrared emission \citep{Joncas}. In summary, the S187 environment is a good target to study the interaction of the ionized, atomic and molecular gas and its impact on star formation. 

In this study, we present new GMRT HI and continuum observational data of the S187 region including two molecular cores (S187~NE \& S187~SE) in the shell around the \Hii\ region that illustrate different phases of the interaction between molecular, atomic, and ionized gases. Section \ref{sct:obs} presents the observational data at 21 cm, radio continuum, molecular lines HCO$^+$(1--0), CS(2--1), $^{12}$CO(1--0), $^{13}$CO(1--0), C$^{18}$O(1--0), and infrared data from WISE. Section  \ref{sct:results} describes the observational results derived from these datasets. More details and the analysis of the observations are presented in Section \ref{sct:analysis}. Also, we discuss our results in Section \ref{sct:discussion}, and at the end we give the conclusions.

\section{Observations and archival data} \label{sct:obs}
\subsection{21~cm line and radio continuum}
The 21-cm data were obtained using the Giant Metrewave Radio Telescope (GMRT, project 26\_012) during 2014 September 4-5. The legacy GMRT backend was configured to the 8.138~kHz (1.72~\kms) resolution and the 4166.7~kHz bandwidth. 3C48, 0102+584, 0217+738, and 3C147 were observed as calibration sources. 3C48 was used as a bandpass and fluxscale calibrator, 3C48, 0102+548, and 0217+738 as phase calibrators. Absorption features were detected towards 3C147, 0217+738, 0102+584. 
The channels affected by absorption were excluded during the calibration. Around 30\% of the data were manually flagged mostly due to radio frequency interference in some baselines. 
The calibration was performed in {\sc CASA} \citep{casa}.
A strong compact radio source \nvssrg\ in the source field was used for phase self-calibration. The continuum image RMS noise was 0.1~mJy{/beam} at the field centre for Briggs-weighted image restoring (beam: 2.28$\times$2.1 arcsec,~PA~-2$^\circ$). {A} UV-tapered continuum image was also produced to achieve the spatial sensitivity with a beam of 8.3$\times$7.0 arcsec,~PA~-10$^\circ$ and 0.7 K RMS.

Two spectral cubes were processed with different restoring beams. One with the Briggs weighted beam (2.7$\times$2.2 arcsec,~PA~-13$^\circ$) and the other with the uv-tapered naturally weighted (8.11$\times$7.28 arcsec,~PA~7.4$^\circ$) beam to achieve sensitivity to the extended structures. We performed uv continuum subtraction for the uv-tapered data and combined it with the WSRT CGPS \Hi\ line survey \citep{cgps} using {\sc CASA sdintimaging} \citep{sdi} to fill the flux loss. {The standard {\sc{clean}} procedure was applied with cutoff limit of 10 mJy/beam.}  The tapered combined image has {a} per-channel RMS noise around 3~K, the Briggs weighted GMRT-only datacube has {a} sensitivity of 1~mJy{/beam} (80~K) RMS noise per channel.

\subsection{Molecular lines}
New HCO$^+$(1--0) observations were carried out in 2021 April with the 20-m telescope of the Onsala Space Observatory (OSO) (project O2020b-02). 
The 3~mm dual-polarization receiver with a SIS mixer was used \citep{Belitsky}. 
{A} Fast Fourier Transform spectrum analyser with a 76~kHz frequency resolution was used. 
The spectrometer gives a velocity resolution of $\sim 0.26$~km\,s$^{-1}$ at 89.2~GHz.
To obtain high-sensitivity data, the signals from two polarizations were co-added during the reduction process, which provided an increase of sensitivity by a factor of $\sqrt{2}$. 
The observations were performed in the frequency switching mode. 
The system noise temperature varied in the range of $\sim 170-290$~K depending on weather conditions. 
The main beam efficiency was $\sim 0.44$ for elevations of $\sim 30^\circ$ at which the source was observed.
The full beam width at half maximum (FWHM) of the OSO-20m telescope at 86~GHz is 43 arcsec. 
Pointing and focus were checked regularly by observations of SiO masers.
Mapping was done with a 20 arcsec grid spacing. 
The line data were processed using the {\sc CLASS} program from the {\sc GILDAS }
package\footnote{http://www.iram.fr/IRAMFR/GILDAS}, the {\sc XS} software\footnote{https://www.chalmers.se/en/researchinfrastructure/oso/radio-astronomy/Pages/software.aspx}, and our original routines.

The CS(2--1) line was observed by Joncas. It was obtained on 1987 November 30 using the 14~m telescope of the Five College Radio Astronomy Observatory (FCRAO) observatory. The data were not previously published. The telescope was equipped with a cooled 3~mm receiver connected to a digital spectrum expander as the backend with resolution of 50~kHz and 256 channels. The weather was clear, and typical system temperature around 400~K. The individual scans were processed using the {\sc IRAF} software \citep{IRAF} and mapped using our scripts based on {\sc astropy} package \citep{astropy}. The spectral resolution is 0.25~\kms\ and the angular resolution is 45 arcsec. The beam efficiency was taken from FCRAO GRS as 0.48 \citep{Jackson}.  We assume the same flux uncertainty of 15\% as in this paper. This data has a larger coverage (1$^h$23$^m$51$^s$.2 +61$^\circ$51$\arcmin$00{\arcsec}9 centre, 10 arcmin square field) than the CS (2--1) observations reported in \citet{Zinchenko2009}. 

Observations of the $^{12}$CO (1--0), $^{13}$CO (1--0) and C$^{18}$O (1--0) lines were kindly provided by Kim Ardvisson and Cris Ardvisson and were done with the 14~m FCRAO telescope, processed by C.~Brunt with the spectral resolution of 0.25~\kms\ and angular resolution of 46 arcsec \citep{Arvidsson}. We also used $^{12}$CO (1--0) data from FCRAO CGPS to study the extended structures around the S187 region. The HCO$^+$ (1--0) data towards \iras\ are from \citet{Zinchenko2009}. They were obtained {at} the Onsala observatory. The spectral resolution is 84~m\,s$^{-1}$ with  {a} 0.4~K RMS noise.

 \subsection{IR data}
 
We used the WISE archive data images to study the dust emission in the region. The 2MASS (H and Ks) \citep{2mass} and WISE (i.e., W1, W2, and W3) photometric data {we utilized} to classify YSOs in the area and Spitzer IRSA, unWISE \citep{unWISE} and WSSA \citep{WSSA} archive data for presentation purposes. We also used {the} SCUBA data archive for a 850~$\mu$m image that was previously described in \cite{Arvidsson}.

\section{Observational results}\label{sct:results}
We present the 1420~MHz continuum and the 21-cm line data in absorption and emission. We discuss the compact features first and then concentrate on the extended structures. We also present molecular gas emission morphology in addition to those published in \citet{Arvidsson}.

\subsection{Radio continuum}
The structure of the 1420~MHz continuum emission is consistent with previous studies \citep{snell,Arvidsson}. Six compact sources are detected in the field, five of them are near the S187 \Hii\ region (Fig. \ref{fig:continuum_spitzer}), four of them are identified in the NVSS catalog.  The fluxes are given in Table~\ref{tab:flux}. 

\begin{figure*}
    \centering
    \begin{minipage}{0.49\textwidth}
        \includegraphics[width=\textwidth]{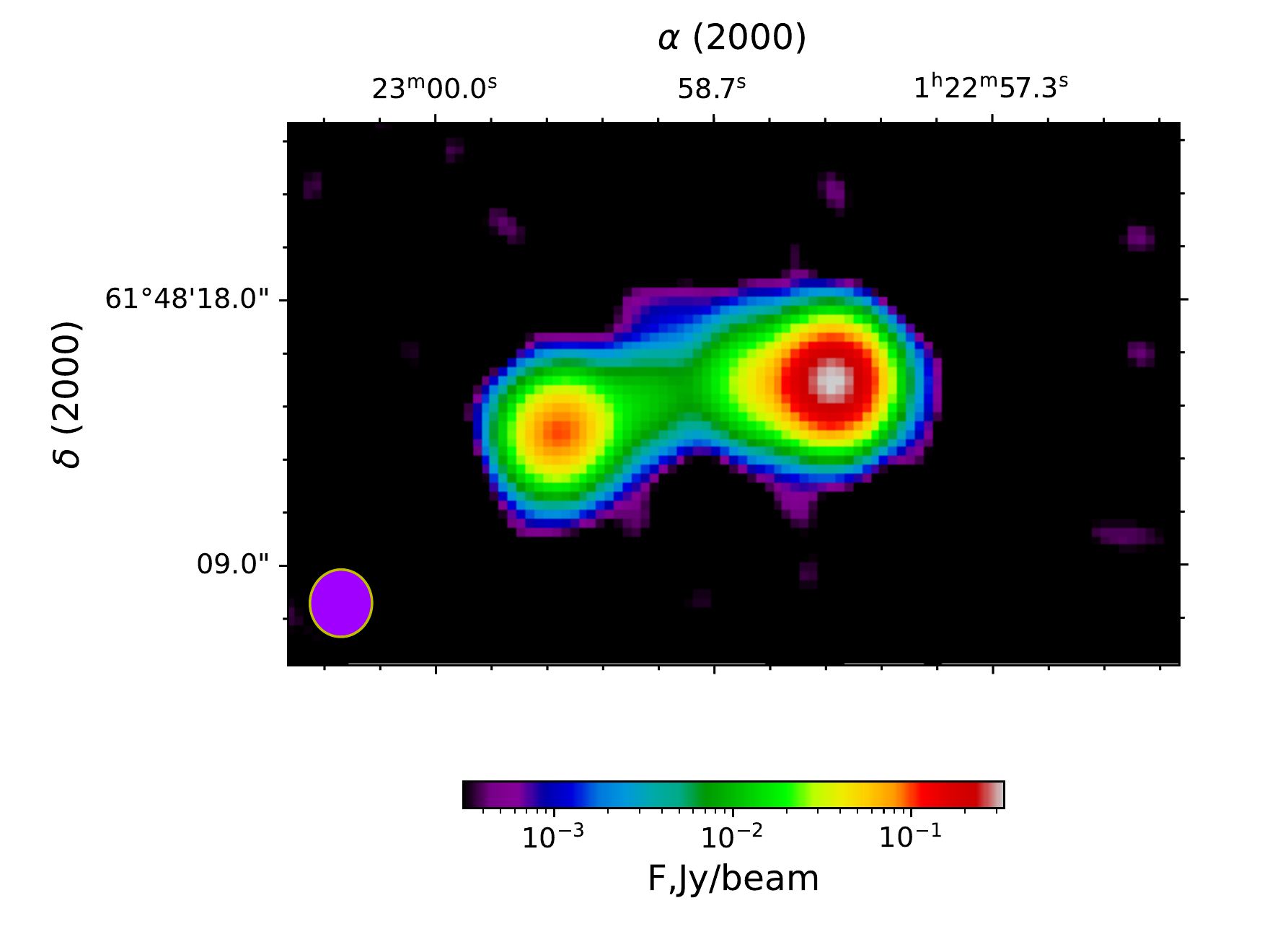}
    	\caption{Closeup view of \nvssrg\ (S187-1 a\&b) in the GMRT 1420~MHz continuum. The beam is shown in the lower left corner. Fluxes are in logarithmic scale.}
    	\label{fig:rg_cont}
    \end{minipage}\hfill
    \begin{minipage}{0.49\textwidth}
        \centering
	\includegraphics[width=\textwidth]{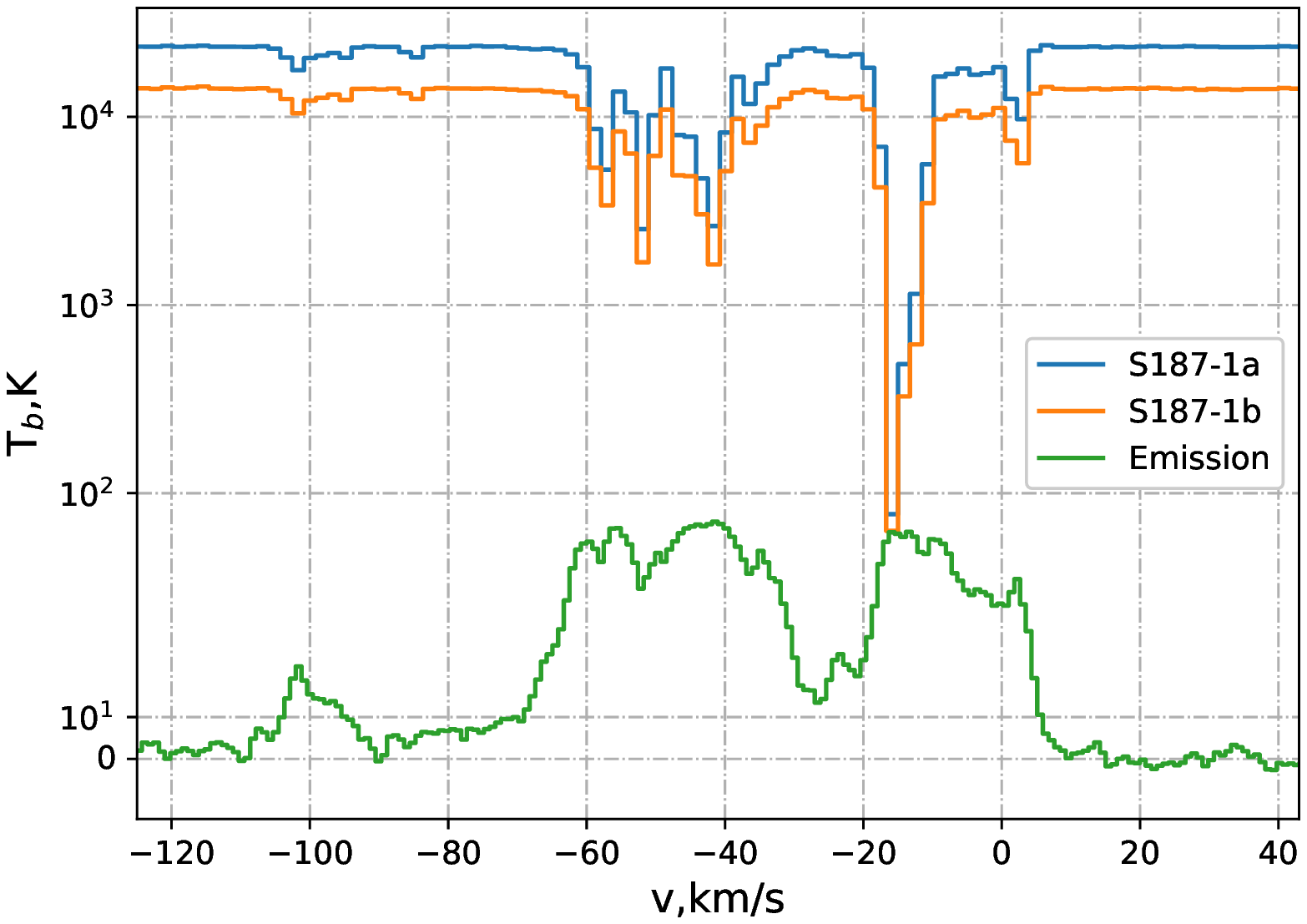}
	\caption{The \Hi\ spectra towards \nvssrg\ (a and b). Emission  {spectra} are extracted from the CGPS survey inside {the} $1^\prime$ {circle around} the galaxy {position}. The absorption profile{s} {are} obtained {from the} GMRT data. {T}he brightness values below 50~K {are} {shown using a} linear {scale} {those} above 50K {with a} logarithmic scale.}
	\label{fig:hi_abs}
    \end{minipage}
\end{figure*}

The parameters were estimated using CASA 2D Gaussian feature fit. The brightest one, \nvssrg\ (S187-1) is located near the S187 \Hii\ region. It is known as a radio galaxy \citep{snell,Arvidsson}. We resolve it in two separate compact sources and an elongated extended weak structure. The compact sources are separated by 10 arcsec  (Fig. \ref{fig:rg_cont}), with brightness of 24000$\pm$120K (S187~1a) and 14300$\pm$120K (Fig.~\ref{fig:hi_abs}). NVSS~012250+61545 (S187~2) is also split into two compact features. A weak point source with a flux of 0.67$\pm$0.14~mJy is detected towards \iras. This source is associated with the Class I protostar reported in \citet{Kang2012}. A water vapor maser is present near \iras\ as demonstrated by \citet{zin98}. This source is a region of high-mass star formation \citep{Kang2012}. To the best of our knowledge, no radio continuum emission was previously reported towards \iras. Thus, there is no data to estimate the spectral index. The continuum point source can represent an UC \Hii\ region. 

S187-1a\&b are brighter than the extended \Hii\ region requesting high quality calibration. {The} UV-tapered continuum image IS shown in Fig.~\ref{fig:continuum_spitzer} in red. The emission associated with S187 is not homogeneous. The shape is close to a ball-like structure overlaid with a curved feature. This feature is bordered {by} 850~$\mu$m SCUBA emission associated with the parental cloud of \iras, as seen in Fig.~\ref{fig:continuum_spitzer}. There are two weak peaks in radio continuum emission, located near {the} SCUBA~850~$\mu$m and WISE~12$\mu$m peaks (Fig.~\ref{fig:continuum_spitzer}). We associate these features with PDRs and call them S187~PDR1 and S187~PDR2, (coord. in Table~\ref{tab:molecular}, see Fig.~\ref{fig:features}). The SCUBA~850~$\mu$m emission is detected near the IR peak{s} {associated with PDR1 and PDR2}. Weak continuum emission surrounds the S187~PDR2 from north and south (Fig.~\ref{fig:continuum_spitzer}{, Fig~\ref{fig:features}, center; , Fig~\ref{fig:features}, right}). Such morphology is an argument for a small, density bounded zone bordered or surrounded by the leaking ionising emission and gas.

\begin{table*}
\caption{Continuum sources properties and parameters estimated using the CASA 2D Gaussian feature fit.}
\label{tab:flux}

\begin{tabular}{c|l|l|l}
\hline
    Source & Coordinates, J2000                                     & Flux at 1420~MHz, mJy  &
    Size, deconvolved: major,minor,p.a., arcsec  \\
    \hline
    S187-1a &  01$^h$22$^m$58$^s$.11304(.0006)     +061$^\circ$48\arcmin15\arcsec.188(.0037)        & 386.4(2.5)    & 1.27(0.02)x0.38(0.063)x91(2.5)\\ 
    S187-1b &  01$^h$22$^m$59$^s$.39627(.002)      +061$^\circ$48\arcmin13\arcsec.52(.013)          & 116.7(0.8)    & 1.4(0.08)x0.35(0.2)x112(4.9)\\
    S187-2a &   01$^h$22$^m$51$^s$.346(.0017)      +061$^\circ$54\arcmin48\arcsec.22(.015)          & 13.7(0.3)     & 1.3(0.1)x0.58(0.16)x144(7.4)\\
    S187-2b &   01$^h$22$^m$50$^s$.91(.02)         +061$^\circ$54\arcmin51\arcsec.93(.14)           & 7.9(0.7)      & 4.42(0.4)x3.4(0.4)x131(22)\\
    \iras   &  01$^h$23$^m$33$^s$.36(.012)          +061$^\circ$48\arcmin48\arcsec.09(.15)   & 0.67(0.14)    & point \\
    NVSS 012355+614404 & 01$^h$23$^m$54$^s$.1(.012) +061$^\circ$43\arcmin57\arcsec.73(.08)   & 1.8(0.24)     & point \\
    S187    & 01$^h$23$^m$04$^s$.8(.1)             +061$^\circ$51\arcmin41\arcsec.2(.4)                      & 1200(20)     & 175.62(3.6)x141.3(3)x162(4)\\
    \hline
    
\end{tabular}
\end{table*}

\begin{table*}
\caption{Molecular and infrared sources sizes estimated using the CASA 2D Gaussian feature fit. The integrated \ceo\ intensity image is marked as $\int$\ceo.}
\label{tab:molecular}

	\begin{tabular}{c|l|l|l}
		\hline
		Source      & Coordinates, J2000  &    Size: maj,min,bpa, arcmin & source\\
		\hline
		S187-NE     &  01$^h$23$^m$29$^s$.9(1.37)     +061$^\circ$54\arcmin28\arcsec.8(8.9)             & 4.1(0.4)x3.65(0.2)x129(7.5) & $\int$\ceo \\ 
		S187-SW     &  01$^h$23$^m$30$^s$.5(1.2)      +061$^\circ$48\arcmin49\arcsec.7(7.0)            & 4.2(0.4)x3.4(0.3)x93(17)   & $\int$\ceo \\
		S187-PDR1   &   01$^h$23$^m$17$^s$.93(.03)      +061$^\circ$53\arcmin10\arcsec.34(.2)           & 3.12(.01)x2.00(.006)x128(0.29) & 12$\mu$m WISE\\
		S187-PDR2   &   01$^h$22$^m$50$^s$.2(.02)         +061$^\circ$52\arcmin28\arcsec.8(.1)          & 2.04(.008)x1.27(.005)x69.4(.3)& 12$\mu$m WISE\\
		HI subbuble &  01$^h$23$^m$39$^s$.8(.1)          +061$^\circ$51\arcmin37\arcsec(1.5)            &  3.9(.1)x4.1(.1)x90& 12$\mu$m WISE\\
		\hline
	\end{tabular}
\end{table*}

\subsection{The \Hi\ data}
Using the combined GMRT+CGPS data we are able to obtain {full-scale} high-resolution datacubes of the atomic hydrogen lines. From the GMRT data only we obtain high-contrast absorption profiles towards S187-1(a\&b). The 21-cm emission and absorption line profiles are shown in Fig.~\ref{fig:hi_abs}. These profiles contain at least 8 spectral features. The emission features at [--5;--20]~\kms\ are located near the S187 \Hii\ region, at the same velocities as molecular gas traced by $^{12}$CO (discussed later). These velocities {argue for an} Orion arm {membership} \citep{Joncas}.

\begin{figure*}
    \begin{minipage}{0.99\linewidth}

	\includegraphics[width=\linewidth]{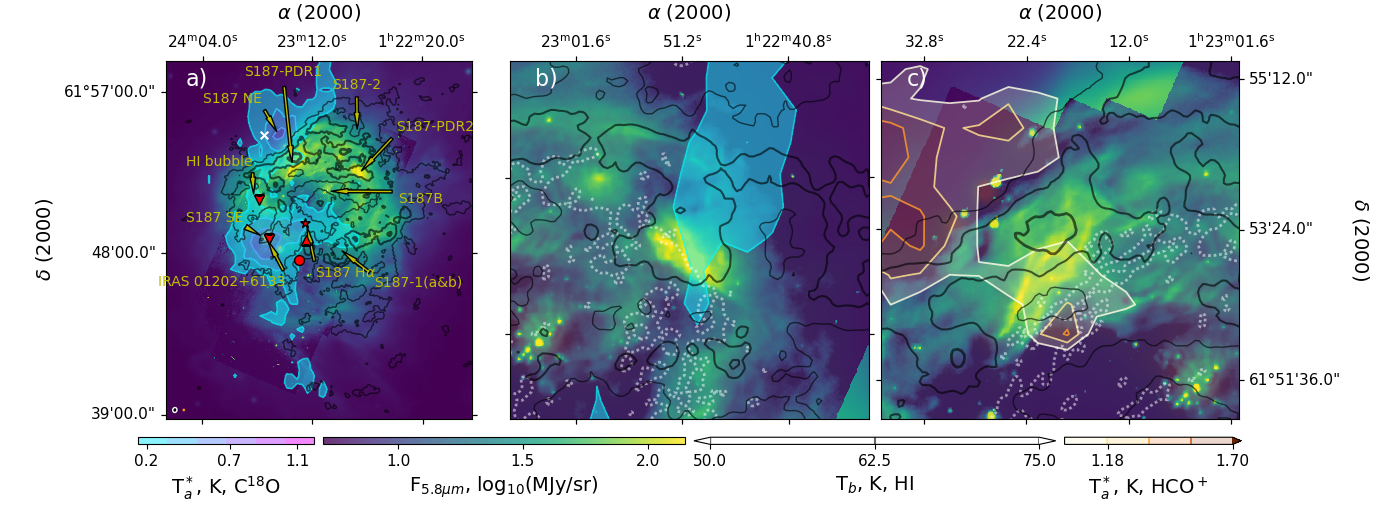}
	\caption{
	{
	\textit{a)} The observed features in the S187 complex. Background image is the Spitzer 5.8$\mu$m image overlaying the WISE WSSA 12$\mu$m image (in green and yellow shades). The \ceo\ average intensity in the [-17.5,-12] km/s range is shown in blue filled contours {while} black contours represent the 21-cm line emission in the same velocity range. The contour width is increasing towards the brighter emission. The beams (46 arcsec, 8 arcsec, and 8 arcsec for the \ceo, \Hi\ and 1420 MHz continuum, respectively) are shown in the lower left corner of the image. {the position of}
	S187~H$\alpha$ is marked as a red star. A white cross marks the HCO$^+$ line intensity peak position. {The} other marks are the same as in Fig~\ref{fig:continuum_spitzer}.
	\textit{b)} 
    	Close-up picture of S187 PDR2. The image scheme is the same as for \textit{a)}, except the velocity range is [-13,-10] km/s. Light gray dotted contours represent 1420 MHz continuum emission at a 7K intensity level.
    	Declination tick marks are placed at +061$^\circ$51\arcmin24\arcsec and +061$^\circ$51\arcmin36\arcsec.
	\textit{c)} 
	    Close-up picture of S187 PDR1. The image scheme is the same as for \textit{b)}, except {the line which is HCO$^+$} the velocity range is [-17.-13] km/s {since the S187 PDR1 central velocity is different}.
	}}
	\label{fig:features}
	\end{minipage}
\end{figure*}

We resolved \nvssrg\ as two compact sources giving the possibility for highly accurate estimation of the optical depth and column density of the absorbing gas (see next section). Other compact sources are too weak to detect absorption between continuum and hydrogen line brightness levels ($\sim$100 K) in our beam.

The 21-cm line {emission from the combined} CGPS+GMRT data are shown {as} channel maps {in} Appendix B (Fig.~\ref{fig:hi_12co_chmap}). The maps are overplotted {with} black contours which show the $^{12}$CO (1-0) emission from \citet{Arvidsson}. The observed structure{s are} in good consistency with those discussed by \citet{Joncas} despite an order of magnitude improv{ement in} resolution (from 60 arcsec to 8 arcsec) {and} a factor of {3} improvement in sensitivity (from 0.7 to 0.{3}~mJy{/beam}) {in restored beam}. The \Hi\ emission towards S187 is not homogeneous, the channel maps {reveal} spatial{ly and} kinematically connected structures up to scales of $\sim$0.6~pc. The features that are spatially related to the \Hii\ region are detected from --20.7 to --3.5~\kms. At the edges of the velocity range, the emission {}is located towards the centre of the \Hii\ region tracing the parts of the {shell walls} oriented with the motion {parallel} to the line of sight. As shown in Fig.~\ref{fig:features}, the integrated \Hi\ line emission correlates well with the {5.8}~$\mu$m {Spitzer} emission. The local peaks of the \Hi\ line are slightly shifted from the infrared data peaks outwards of centre of the \Hii\ region. The emission region has a shape close to a circle with inhomogeneous borders except {for} the north-east direction, where the border is curved inside.

\subsection{Molecular lines}

The shell around S187 has a molecular counterpart \citep{Joncas,Zinchenko2009,Arvidsson}. It consists of dense regions with molecular cores and more diffuse, surrounding gas generally related to the ionized region envelope. We used the CS(2--1) and HCO$^+$(1--0) observations as tracers of the dense regions. The molecular gas emission in the CO isotopologue lines are presented AS channel maps (Figures in appendix B, \ref{fig:hi_c18o_chmap}, \ref{fig:hi_12co_chmap}, \ref{fig:hi_13co_chmap}). The $^{12}$CO emission lines are detected {all over} the field. The $^{13}$CO emission region is more extended than the \Hi\ associated with S187. The \ceo\ emission is usually optically thin, tracing the column density of the molecular gas. We associate it with the dense cores in the molecular shell at different stages of evolution ({d{i}scussed below}). There are two peaks detected at different velocities, hereafter called S187~SE, the parental cloud associated with \iras\ and S187~NE. S187~NE spatially and kinematically correlates with the $^{12}$CO peak while S187~SE does not. At the velocity of S187~SE ($\sim$ --15 \kms) the $^{12}$CO line is mostly homogeneous (Fig. \ref{fig:hi_12co_chmap}). The bright peaks of the \ceo\ emission strongly anti-correlate with the \Hi\ data, {most} strongly towards S187~SE. The molecular emission elongates towards {the} S187 exciting star cluster, a small peak of the \Hi\ emission is detected towards the centre of the \Hii\ region. No 12$\mu$m WISE feature is detected towards or near S187~SE. This absence can be {a} projection effect on the line of sight meaning that the molecular gas is not related to the S187 shell, but the anti-correlation between \Hi\ and \ceo\ suggests that the molecular gas of S187~SE is embedded into the atomic medium. No \Hi\ emission is detected at the position and velocity of S187~SE \ceo\ and HCO$^+$ peak{s}. No cometary $^{12}$CO molecular emission is detected around the \ceo\ peak. The peak of \Hi\ gas, at the -11~\kms\, is located towards the molecular emission peak near the S187-H$\alpha$ object. 

\begin{figure}
	\includegraphics[width=\linewidth]{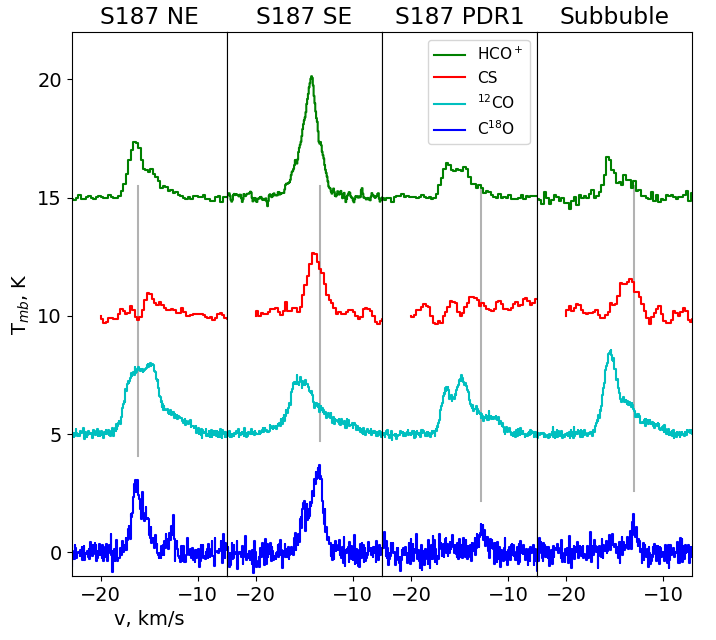}
	\caption{The \ceo(1--0), $^{12}$CO(1--0), CS(2--1) and HCO$^+$(1--0) spectra towards some positions of the S187 shell. The $^{12}$CO(1--0) intensity {is} divided by {a} factor of 7. Vertical line{s} mark {the} \ceo\ peak intensity.}
	\label{fig:specs}
\end{figure}

The CS line properties (Fig. \ref{fig:specs}) are consistent with those in \citet{Zinchenko2009}. In addition we detect an extended feature with a line brightness of $\sim$0.6$\pm$0.2~K and central velocity of --15 \kms. The feature {extends} to the north of \iras.  The CS and HCO$^+$ peaks are shifted from the \ceo\ peak outwards of S187B. The HCO$^+$ observations towards S187~NE have similar properties. Two features also exist. The line peak is centred at 1$^h$23$^m$34$^s$.7 +61$^\circ$54$\arcmin$38$\arcsec$.8 with maximum intensity of 2.8$\pm$0.2~K and velocity of --16$\pm$0.3 \kms\ and a $\sim$2 \kms\ line FWHM. (Fig. \ref{fig:features}, a)) The extended feature has a $\sim$3 \kms\ FWHM centered at --15$\pm$0.3 \kms\ and peak intensity of 1.7$\pm$0.25 K. The extended feature is located towards the S187~PDR1. The profile has non-gaussian features. The line profiles are shown in Fig. \ref{fig:specs}.

\section{Physical properties estimation} \label{sct:analysis}
\subsection{The properties of the \Hi\ gas}
The presence of two closely-located strong background sources (S187-1a\&b) gives {the} possibility for accurate measurement of the absorption profile of the cold neutral hydrogen gas. {We used GMRT-only data (beam: 2.7$\times$2.2 arcsec,~PA~-13$^\circ$) for optical depth profile {extraction}.} Using the emission line from the CGPS survey we are able to estimate the spin temperature of the neutral hydrogen. We assume that the properties of the background galactic atomic gas does not vary significantly across the GMRT field.

\begin{figure}
	\includegraphics[width=\linewidth]{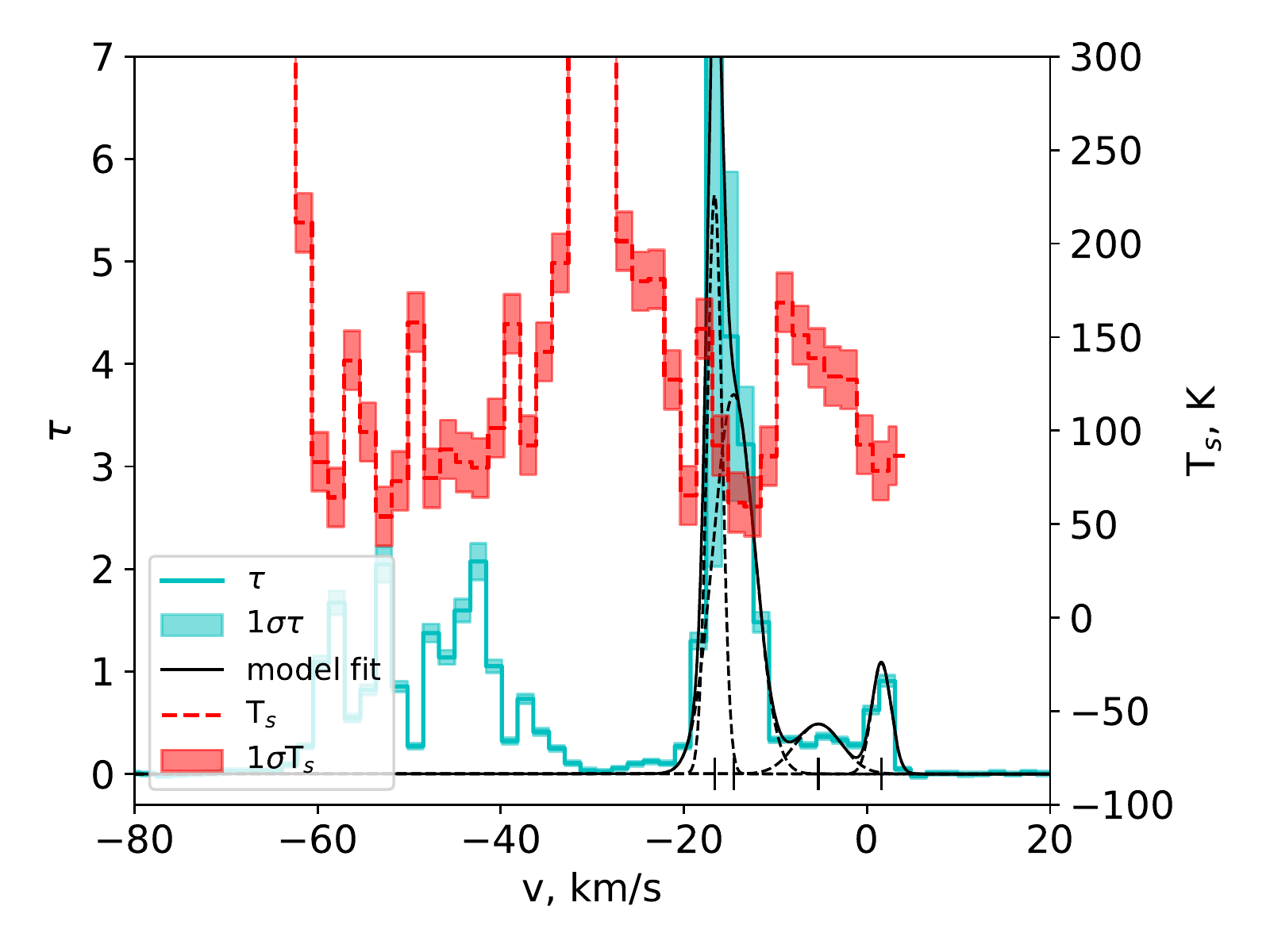}
	\caption{Optical depth and spin temperature profiles towards S187-1(a\&b). {The flat-layer model fit is shown as a black solid line and the individual layers emission profiles are shown as dashed lines. The velocity of each layer is shown by a vertical mark.}}
	\label{fig:tau_spine}
\end{figure}

We estimate the optical depth towards S187-1(a\&b) from {the} two \Hi\ absorption spectra:
\begin{equation}
\tau(v)=-\ln \left[ \frac{T_{br}^{a}(v)-T_{br}^{b}(v)}{T_{c}^{a}-T_{c}^{b}}\right],
\label{eq:tau}
\end{equation}

where $T^a_{br}(v)$ and $T^b_{br}(v)$ are the observed line temperatures towards the S187-1 components $a$ and $b$ shown in Fig.~\ref{fig:hi_abs}. $T^a_c$ and $T^b_c$ are the continuum brightness temperatures of the components derived from the same spectra. We suppose the same filling factor for both components separated by 9.5~arcsec~(0.05 pc) and the same optical depth for the absorbing gas towards both sources. The components $a$ and $b$ have nearly the same size (Table. \ref{tab:flux}). {The sizes estimated as 1$\times$3 \arcsec and 1$\times$2 \arcsec (5 \& 15 GHz) for $a$ and $b$, respectively \citep{snell}.}  {With} this estimation we also exclude calibration errors between different instruments and missing flux related to the absorption-emission optical depth estimation \citep{Nguyen}, where absorption spectra are acquired by an interferometer and emission spectra by single dish. The spin temperature is calculated using emission spectra averaged near S187-1(a\&b) from the CGPS survey observations over a 1$'$ region excluding the S187-1 area:
$$T_s=T_{em}/(1-e^{-\tau(v)}),$$
where $T_{em}$ is the line temperature from the CGPS survey (see Fig.~\ref{fig:hi_abs}) and $\tau$ is taken from Eq.~\ref{eq:tau}. The resulting profiles are shown in Fig.~\ref{fig:tau_spine}. {The high uncertainty of {the} optical depth estimation at -17.5 \kms\ may be caused by two factors. First, the presence of extended emission {at that} channel increases the RMS and {thus} the uncertainty of the estimation. Second, S187-1~a and S187-1~b have {similar} intensities due to the high optical depth which leads to A smaller value of the numerator in the expression \ref{eq:tau} and {thus} higher uncertainties.} It is worth noting that $T_s$ and $\tau$ {estimations} are the line of sight ensemble-averaged values. {We do not take into account a possible beam dilution that may increase the spin temperature value.}

To estimate the properties of the atomic hydrogen layer in the entire S187 shell we {fit} a flat-layers model {to the emission spectra from the combined data}. The brightness temperature $T_b$ emitted by layer $i$ is 
\begin{equation}
\label{eq:br}
T^i_b(v)=T^i_s(1-e^{-\tau_i(v)})+T^{i-1}_b(v)e^{-\tau_i(v)},
\end{equation}
where 
\begin{equation}
    \tau_i(v) = \tau_i \times exp(-(v-\Delta v_i)^2/(2\sigma_i)),
    \label{eq:tau_fit}
\end{equation}
is the optical depth profile of layer $i$, $\tau_i$ is the peak optical depth of layer $i$, $\Delta v_i$ is the Doppler velocity and $\sigma_i$ is the {Doppler} width of the optical depth profile of the same layer. The total number of layers is defined by the number of components in the Gaussian decomposition of the optical depth profile derived from the absorption spectra and was equal to {4}. The velocity of the layer decreases with $i$. We assume that the outer layers represent outer parts of the Galaxy. 

We analysed the {combined CGPS+GMRT } data using the algorithm described in Appendix \ref{appendix:fitting} {which combines iterative least squares fitting whose initial conditions {are} estimated by k-Nearest Neighbors (kNN)}. Based on the analysis of the spectral profiles in different positions of the map we constrained some parameters of the different layers. The component at the velocity of $\sim$ 3~\kms\ does not vary significantly and their parameters are fixed. We also fixed the velocity, 
the Doppler width and the spin temperature of the component at $\sim$ -5~\kms. These components are probably related to other galactic structures. 
Summarizing, nine parameters are varied,  ($T^i_s$,$\tau_i$,$\Delta v_i$,$\sigma_i$) 
for each component and $\tau_i$ for one component. 

Using the constraints described above we fitted the model to the different positions. The results converged in three passes. The optimal number of the nearest neighbors is estimated as 5. 

\begin{figure}
	\includegraphics[width=\linewidth]{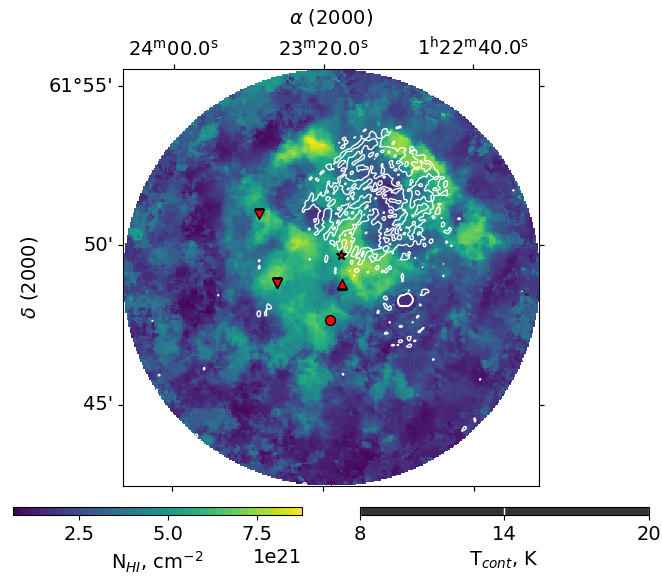}
	\caption{The atomic gas column density map towards S187. The white contours represent continuum 1420 MHz emission. The marks are the same as in Fig~\ref{fig:features}.}
	\label{fig:hi_colden}
\end{figure}

As described in the next section, the \Hi\ gas has the internal kinematics of {an} expanding bubble, so the assumption of a Gaussian line width in this object is far from ideal. We estimate the column density of the atomic layer from the following equation using the fit results and the kNN regression:

\begin{equation}
    N_{HI}=1.823\times10^{18}cm^{-2} \int\frac{\tau T_{b}(v)}{[1-exp(-\tau(v))]}dv,
    \label{eq:nhi}
\end{equation} 
from \cite{Saha}. Here $T_b$ is the brightness temperature of the model component and $\tau$ is from Eq.~(\ref{eq:tau_fit}). The column density map of the atomic gas related to S187 is shown in Fig.~\ref{fig:hi_colden}. {The atomic gas associated with S187 is surrounded by ambient gas. We subtract the mean column density of the ambient gas and calculate the mass of the \Hi\ gas related to S187 as 260$\pm$56~M$_\odot$. The ambient gas column density is estimated by the using the flat-layer model fit in the outer regions where no $^{13}$CO and 12 $\mu$m WISE emission was detected.} Using the model fit and assuming the temperature broadening we estimated the upper limit of the kinetic temperature of the atomic gas $T_{kin}<1000$~K. In the case of thermalization, the \Hi\ kinetic temperature should be equal to the spin temperature. The median temperature over the S187 shell temperature is 50~K. The peak of the spin temperature is located towards S187~PDR1 with a value of 110~K. The FWHM of the PV diagram size is $\sim$10 arcmin which we adopt as the \Hi\ shell size. 

The HI column density is {well-correlated} with the bright 12 $\mu$m WISE emission, which is tracing hot dust and polycyclic aromatic hydrocarbons (PAH) {in} PDRs, but with significant deviation from {a} linear regression. Negative correlation with the 12 $\mu$m WISE emission is present in the [0,$\sim$0.0015] Jy{/sr} WISE flux range. The related points are located at the periphery and probably caused by {a fraction of the atomic gas not being} related to S187. The determination coefficient R$^2$=0.26 is relatively low and caused by non-systematic deviation from the linear regression. This deviation exceeds the estimated uncertainties and may trace the different excitation and PAH generation conditions. {It is possible that} not all atomic gas is produced by photo-dissociation of the S187 molecular shell.

\subsubsection{Atomic gas clumps properties}
\label{sct:fell}

\begin{figure}
        \includegraphics[width=\linewidth]{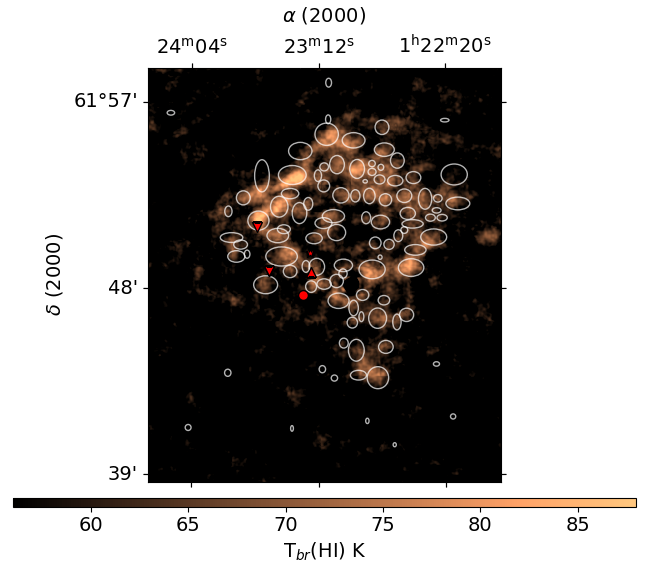}
    	\caption{{Spatial} distribution of {the \Hi\ fragments} identified by {the} CUPID Fellwalker algorithm. The {ellipses} represent clump sizes (standard deviation around centroid). {The \Hi\ peak intensity is plotted in the background. The marks are the same as in Fig~\ref{fig:features}.}}
    	\label{fig:clumpfind}
\end{figure}

We analysed the \Hi\ datacube using the {\sc Fellwalker} algorithm \citep{cupid}. The cutoff limit was set to 7 times {the} RMS and the [-19,-5] \kms\ range is used. The results are shown in Fig.~\ref{fig:clumpfind} and in Fig~\ref{fig:hi_12co_chmap}. We detect $\sim$100 artefacts associated with the S187 atomic shell. We assume that the detected artefacts are the inhomogeneities of the atomic layer or fragments. Most of the fragments are concentrated at the shell periphery and correlate well with the 4.6 and 12~$\mu$m WISE emission. Groups of small ones are located near the S187~PDR1, S187~PDR2 and S187~SE borders. 5\% of the fragments are smaller than the double beam size, most of the fragments are resolved. 

\begin{figure}
    \centering

    	\includegraphics[width=\linewidth]{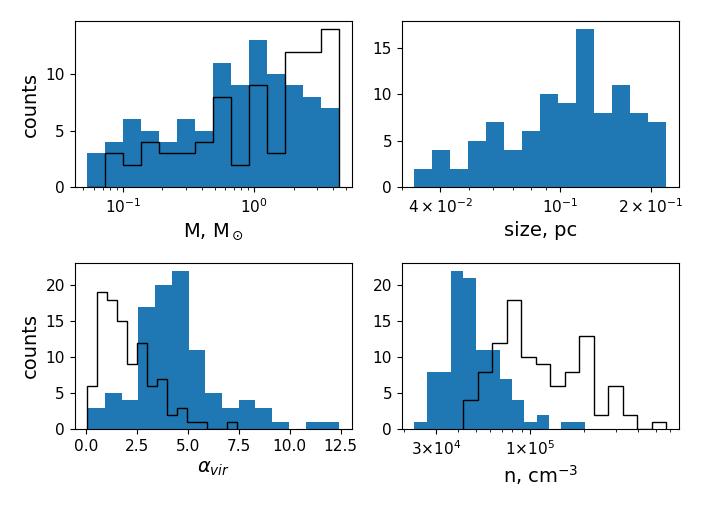}
	    	\caption{
	    	The properties of the fragments.\textit{ Upper left:} Individual fragment atomic gas mass histogram. Upper right: The fragment {linear }sizes histogram. \textit{Lower left:} Virial parameters of the fragments. \textit{Lower right:} the individual fragment volume density  histogram. {The filled bins correspond to the optically thin conditions and the unfilled ones are based on the flat-layer model.} The bin {step} has a logarithmic base except for the lower left plot.
	    	}
    	\label{fig:clump_hist}

\end{figure}

\begin{figure*}
    \centering

    	\includegraphics[width=\linewidth]{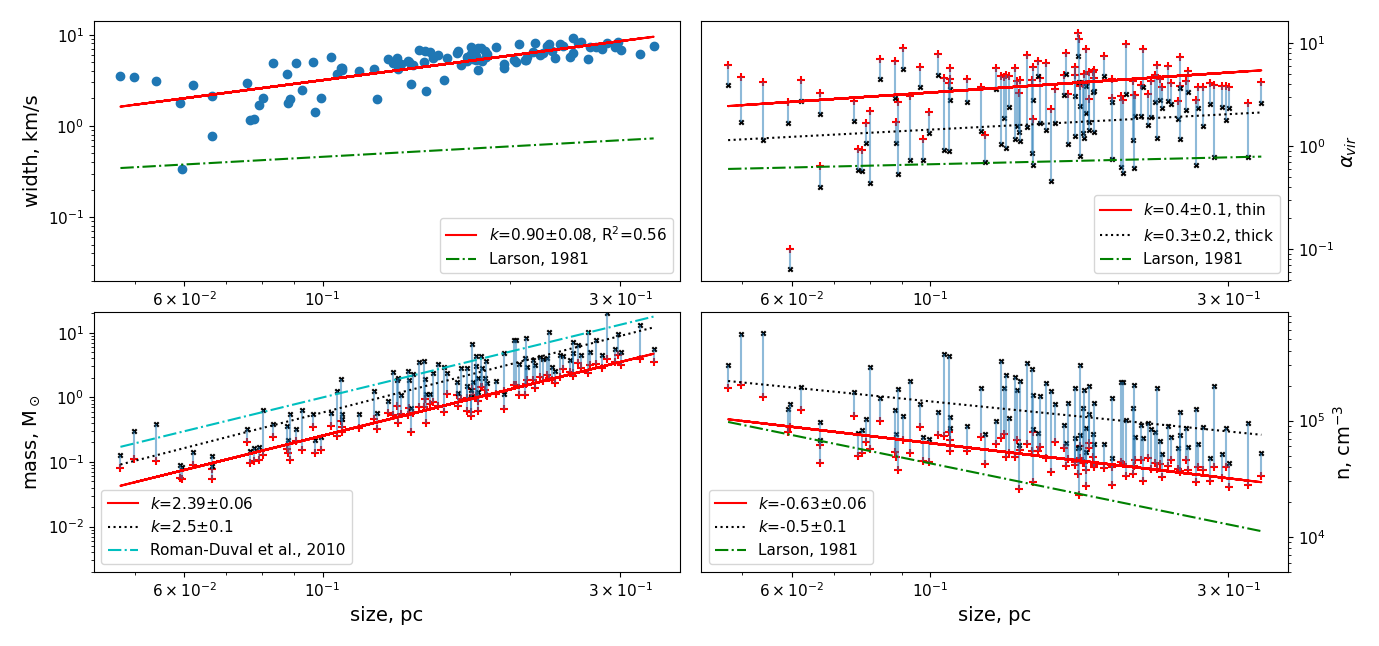}
	    	\caption{
	    	The correlations between fragment parameters. \textit{{Top} left:} Linear size versus {line width}. \textit{Top right:} {Virial parameters} {versus} linear size scatter. \textit{Lower left:} {M}ass versus linear size. {\textit{Lower right:}} Volume density versus linear size. {The red pluses represent the values estimated in the optically thin limit and the black crosses represent the values estimated by using optical depth values from the flat-layer model fit results.} {The best fits are shown as red lines {(thin) and dotted black line ($\tau$ from FLM)}. The same dependencies from \citet{fracpow, Larson81} are shown by dash-dotted lines. }
	    	}
    	\label{fig:clump_correlations}

\end{figure*}

We estimate masses from the column density ($N_{HI}=1.823\times10^{18} cm^{-2}\int T_b dV)${)} and the geometric sizes of the fragments under the assumption of {low} optical depth \citep{willson}. {The fraction of {t}he optical depth concentrated in the fragments compared to into the interfragment medium is unknown so we consider it as the lower limit mass estimation}.  {The fragment size along the line of sight is unknown and is assumed to be the mean value of sizes in the plane of sky.} The results are shown in Fig.~\ref{fig:clump_hist}. The mean mass of the fragment is 1.14 M$_\odot$. The total mass of the fragments {estimated as} $\sim$114~M$_\odot$. {Alternatively, the upper limit of fragment masses can be estimated under the  assumption of the same optical depth as for the flat-layer model. This assumption gives nearly the same total mass of the fragments as for the entire shell ($\sim$240~M$_\odot$). The actual fragment masses should lay between these values.} Typical volume density is around $\sim$5.4$\times$10$^4$ cm$^{-3}$ {for optically thin limit}. Detected fragment sizes vary from 0.03~pc to 0.23~pc. The size is estimated as standard deviation of individual pixels calculated at 1.4~kpc distance, weighted by intensity. Our data indicate a tight (R$^2$=0.94) power-law correlation between mass estimation and fragment size ({Fig. }\ref{fig:clump_correlations}). The power-law index for the mass-size relation is $k=2.39\pm0.06$.



\subsection{The molecular gas properties}

We analysed the $^{13}$CO and \ceo\ data from \cite{Arvidsson}{. The data has} the same spatial and velocity {coverage as \Hi\ combined data.}

The CO isotopologues column density ($N_{\mathrm{CO}}$) is estimated using Eq. 90 from \citet{Mangum15}. We assumed the same excitation temperature for the $^{12}$CO, $^{13}$CO and \ceo\ lines and estimated it from $^{12}$CO data using Eq. (15.30) from \citet{willson}. No $^{12}$CO emission associated with S187~SE {is} detected. {For the S187~SE core,}  the excitation temperature is assumed as 9.5~K using the kinetic temperature estimation from \citet{Kang2012}. The $^{13}$CO/$^{12}$CO=0.0145 abundance ratio is calculated for the S187 galactocentric distance of 9.27~kpc taking into account the galactic gradient of this ratio \citep{Pineda}. The CO to H$_2$ abundance ratio (7.7$\times$10$^{-5}$~for~S187) is calculated {using the} gal{act}ocentric distance \citep{Fontani}. {We estimated the s}hell mass using the $^{13}$CO data as 6400 M$_\odot${. Such estimate }is close to the {value from} \citet{Arvidsson} {derived from} the same data.

We estimated virial parameters {for} the S187~SE and S187~NE cores  {using} $\alpha_{vir}=1.2 \sigma^2 \sqrt{S_{eff}}/(10^{-3} M)${,} where $M$ is the core mass in $M_\odot$, $\sigma$ in \kms\ is calculated using the second moment of the $^{13}CO$ line, $S_{eff}=\int{T_b}ds /T_b$ is the core area in pc$^2$ calculated from the $^{13}$CO $T_b$ image, $ds$ is the elementary area at a distance of 1.4 kpc \citep{Kauffmann}. The $\sigma$ is $\sim$1.4 and $\sim$0.9 \kms, the mass is $\sim$1200 and $\sim$900 M$_\odot$ for S187~SE and S187~NE, respectively. The $\alpha_{vir}$ is estimated as 2.75 and 1.4 for S187~SE and S187~NE, respectively. The core masses are close to each other, the difference between virial parameters is caused by different line widths, that {is} probably caused by non-thermal motions in the core{s}.

\subsection{Dust properties from the Akari-FIS \& WISE data}
We used data from the WISE and AKARI telescopes. The analysis is based on the 3.36, 4.6, 12.08, and 22.19~$\mu$m images from the WISE survey \citep{Wise} and on 65, 90, 140, and 160~$\mu$m from the AKARI survey \citep{Akari}. Our aim was to identify different dust fractions with molecular and atomic spatial-kinematic features in the shell using the colours at different IR wavelengths. Spizer data are also available towards S187, but it has poor coverage and quality. There is no Hershel observations for S187. For the analysis the WISE and AKARI-FIS data are used instead.

\begin{figure}
    \centering
    	\includegraphics[width=\linewidth]{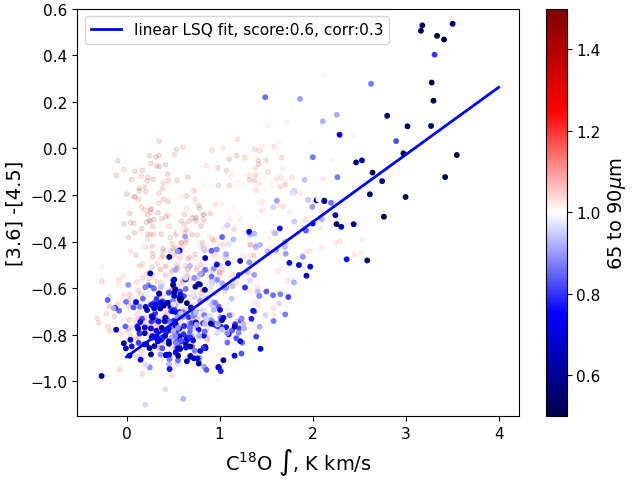}
            \caption{Scatter plot of the {[3.4] $-$ [4.6]}~WISE {colour} vs the \ceo\ zero moment. Linear regression is estimated {by} Huber regression from \citet{scikit-learn}. {Dot} colour indicates the 65~$\mu$m to 90~$\mu$m flux ratio value. Dots outside of the colorbar range and {the} \ceo\ sensitivity limit are plotted as semi-transparent and are excluded from fitting.}
    	\label{fig:ircorr}
\end{figure}

Prior to the analysis of the IR images of S187 and the calculation of the flux ratios, we cleaned the archival IR maps of point sources. {They } were removed from the images using an automatic cleaning procedure. The search and removal of point sources {were} carried out in three subsequent stages depending on the brightness of the stars. More information can be found in \citet{Topchieva2017a}. {T}he data were {then} convolved with a kernel as in \citet{Aniano} and reduced to the same pixel sizes and flux units.

\begin{figure}
	\includegraphics[width=\linewidth]{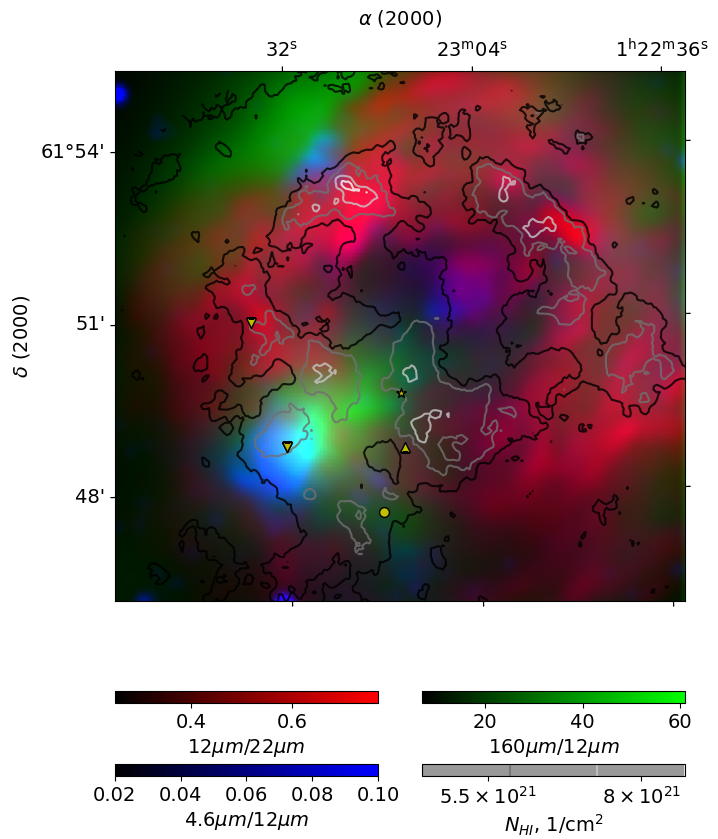}
	\caption{The S187 shell in the IR continuum. The WISE WSSA 12~$\mu$m image is in red. The {\Hi\ column density} is shown in contours. {The marks are the same as in Fig~\ref{fig:features}.}}
	\label{fig:ir_colors}
\end{figure}

The maps {(Fig.~\ref{fig:ir_colors})} are plotted for different ratios of infrared wavelengths. At shorter wavelengths, the emission presumably comes from small grains, while emission in the 70 and 160~$\mu$m bands comes from relatively large grains \citep{Draine}. The emission at 12.08 and 22.19~$\mu$m may come from both stochastically heated small and hot large dust grains and PAH \citep{Pavlyuchenkov} {PAH in 12$\mu$m usually prevail{s} \citep{Reach2}}. The 65 to 90~$\mu$m ratio is used {as a} large{-}grain temperature tracer \citep{Draine}.   

Unlike \citet{Reach}, 
we found that the {the 12$\mu$m to 22$\mu$m ratio} varies from point to point from {0.2} to 0.74 {(See Fig.~\ref{fig:ir_colors})}. The {ratio} correlates with the {\Hi\ column density}. {The ratio variation can be associated with PAH/grains differentiation and processing \citep{Reach}. The ratio peaks are slightly shifted from \Hi\ column density peaks implying that PAH and \Hi\ are effectively generated on slightly different conditions}. 

{We also analysed the [3.4]-[4.6] WISE colour and found that it is associated to the \ceo\ integrated intensity. } The {colour} is plotted in Fig.~\ref{fig:ircorr}. We also plot the 65~$\mu$m to 90~$\mu$m {ratio} from AKARI data in colour. {The colour correlate the \ceo\ intensity under the linear regression. The determination coefficient $R^2$=0.7 for the regression line.} The 90$\mu$m flux dominating region is closer to {the} regression line than {the} 65$\mu$m dominat{ed region}. The 90 $\mu$m emission is usually related to large dust grains \citep{Draine} protected by dense gas represented by \ceo\ emission.

\subsection{Young stellar population}
\begin{figure}
	\includegraphics[width=\linewidth]{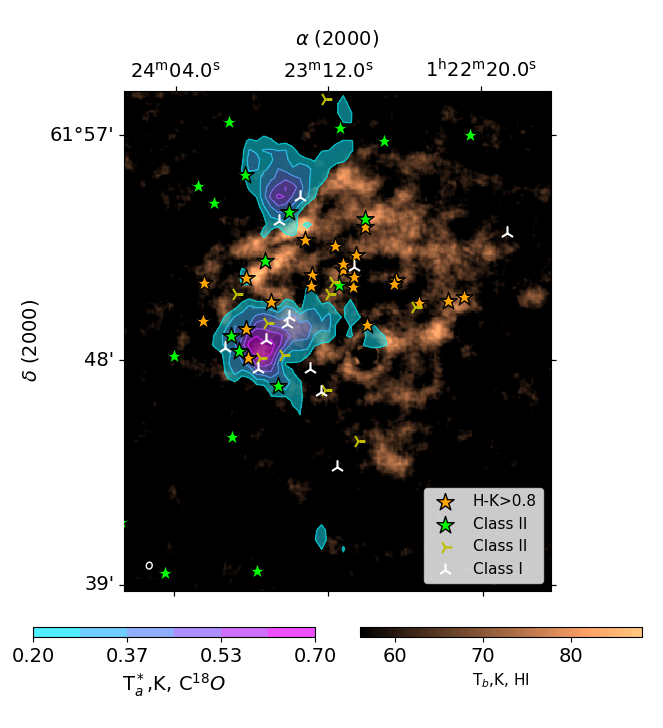}
	\caption{The distribution of the YSOs near S187. The background image represents the \Hi\ averaged emission in {the} [-17.5,-12] {\kms} {range}. The blue contours {represent} the \ceo\ emission {averaged} {for the} same velocity range.}
	\label{fig:yso}
\end{figure}

In order to identify infrared-excess sources or young stellar objects (YSOs) in our selected target area, we used the 2MASS (H and Ks) and the WISE (i.e., W1, W2, and W3) photometric data. \citet{Koenig} presented several schemes to identify YSOs, and two such schemes are utilized in this paper. We used one scheme based on the H$-$K$_{s}$ and W1$-$W2 colo{u}r conditions, and the second one using the W1$-$W2 and W2$-$W3 colo{u}rs. The results are shown in Fig. \ref{fig:yso}.

\section{Discussion} \label{sct:discussion}
Our data trace almost all aspects of the high mass star formation: the \Hii\ region with the ionizing cluster, {the} PDR regions {with} the {photodissociated} \Hi\ layer, the population of young stellar objects, and at least two molecular cores that could be at different {stages of } evolution. In Section~\ref{sct:large_scale} we discuss{ed} large-scale features that may be related to {the origin of the} S187 {complex}. In Section~\ref{sct:shell} we discuss{ed} the general distribution of the gas and YSOs around {and in} the \Hii\ region. In Section~\ref{sct:details} we discuss{ed the} {inner} spatial-kinematic structure {of} the shell{.}

\begin{figure*}
    \centering
    \begin{minipage}[t]{0.45\textwidth}
        \includegraphics[height=1.4\textwidth]{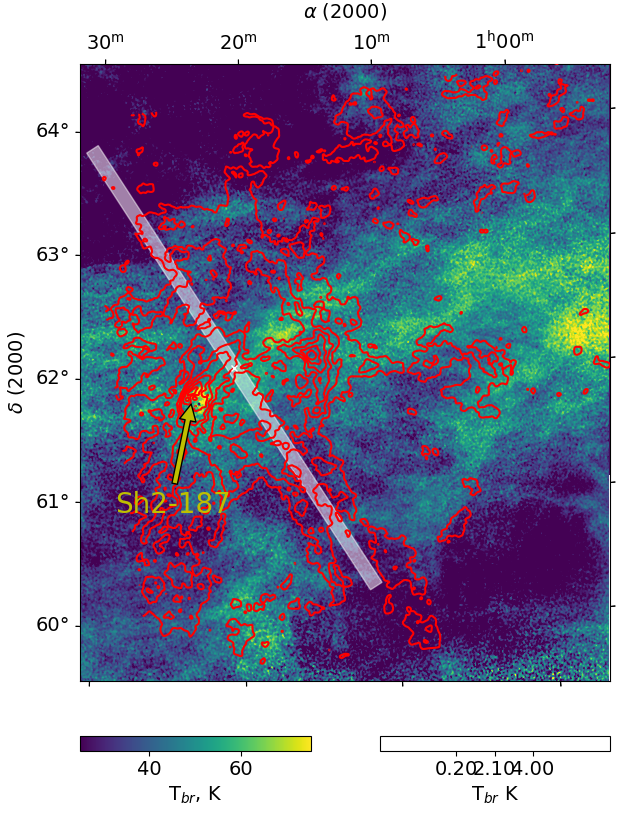}
    	\caption{Map of the S187 surroundings. {The} averaged {CGPS} \Hi\ emission ( [--14~to --17~\kms] range) is shown in the background. It is overlaid with the $^{12}$CO emission {averaged} in the --5 to --17~\kms\ interval. The rectangle indicates the location of the PV diagram shown in Fig.~\ref{fig:snr_pv}. {The crosses correspond to 0 and 2.2 degrees position offsets.}}
    	\label{fig:snr_im}
    \end{minipage}\hfill
    \begin{minipage}[t]{0.45\textwidth}
        \centering
	\includegraphics[height=1.4\textwidth]{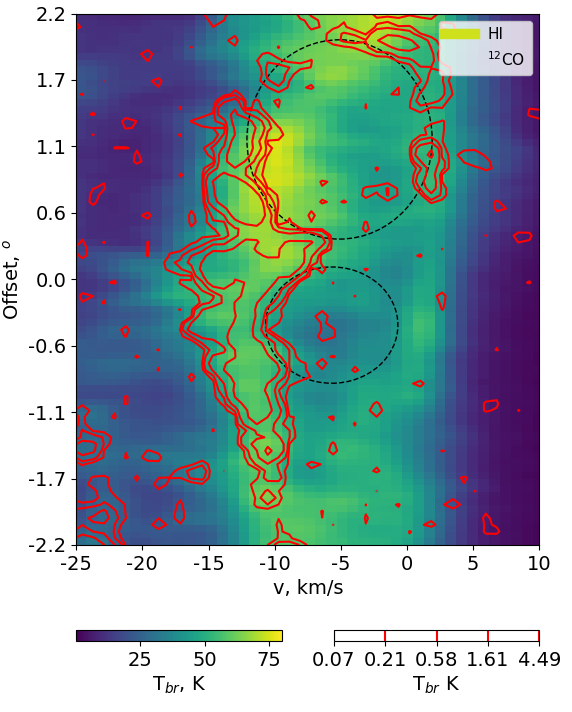}
	\caption{The position-velocity diagram of the atomic and molecular gas near the S187 region. {The PV slice is shown on Fig.~\ref{fig:snr_im}.} Black ellipses indicate the probable SNR locations.}
    	\label{fig:snr_pv}
    \end{minipage}
\end{figure*}

\subsection{Large-scale features}\label{sct:large_scale}

{One of the interesting feature that is probably not related to S187 is the existence of} narrow absorption and emission HI features at $\sim${5}~\kms\ (Fig. \ref{fig:hi_abs} {and \ref{fig:tau_spine}}). {They} are in principle "forbidden" in this quadrant by the rotation curve of the Galaxy \citep{Joncas}. The existence of this feature is interesting but its detailed analysis lies beyond the scope of this paper. Such features are usually explained by local streaming motions near the Sun caused by the Galactic density wave \citep{Burton}. Another possible explanation can be large-scale kinematics caused by the expansion of a supernova remnant \citep{snova}. The picture of the surrounding medium is shown in Fig. \ref{fig:snr_im} and is taken from the CGPS \Hi\ and $^{12}$CO (1--0) surveys \citep{cgps}. A position-velocity diagram IS shown in Fig.~\ref{fig:snr_pv}{. Two nearly circular structures traced in \Hi\ and $^{12}$CO are detected. Such PV {features} can be formed} by the expansion of {energetic} bubbles{. The feature at $\sim$5~\kms\ is probably part of {such a} bubble. The "unusual" velocity can be {explain}ed {by} the {"snow-plow" effect often accompanying such energetic phenomena.}}

We {propose} that these bubbles {are related to} supernova {events}. The northern one is probably caused by SNR G127.1+0.5 \citep{LeahySNR}, and the southern one is probably related to the pulsar B0105+65 and the {G126.1-0.8} shell \citep{snova}. 
We estimate the dynamical age {from} $t_{dyn}=\alpha R/V_{exp}$. $\alpha$ = 0.25 for a radiative-driven expansion and $\alpha$ = 0.6 for a shell created by the action of stellar winds. {T}he inner radius of the atomic part of the shell {is $R$} and $V_{exp}$ is the expansion velocity \citep{snova}. The northern bubble age is t$\sim$3.6$\times$10$^5$~yr (radiatively driven) 
 {or} t$\sim$8.8$\times$10$^5$~yr (wind driven expansion) \citep{snova}. The southern bubble age is t$\sim$3.0$\times$10$^5$~yr {or} t=7.15$\times$10$^5$~yr for {a} radiative {or} wind driven expansion respectively, which gives relatively {the} same age for both SNR.

\begin{figure}

	\includegraphics[width=\linewidth]{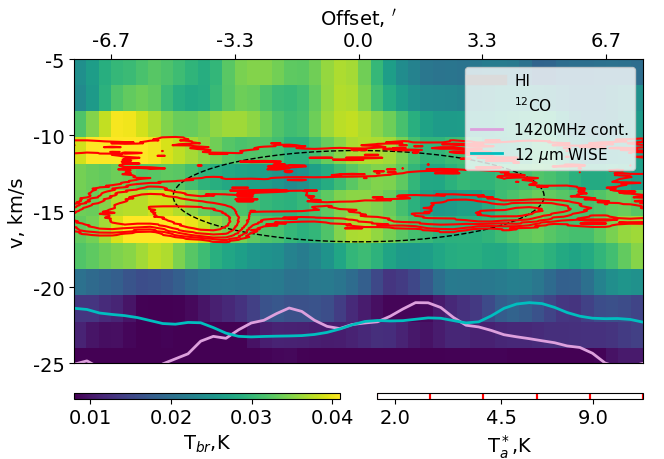}
	\caption{The position-velocity diagram of the atomic and molecular gas. Normalized IR and radio continuum profiles are plotted at the bottom. The ellipse represents {inner border of} the radially expanding shell. The PV {slice} is shown in Fig. \ref{fig:pv_maps} as the red box.}
	\label{fig:shell_pv}
\end{figure}

\begin{figure*}
    \begin{minipage}[t]{0.48\textwidth}
    	\includegraphics[width=\linewidth]{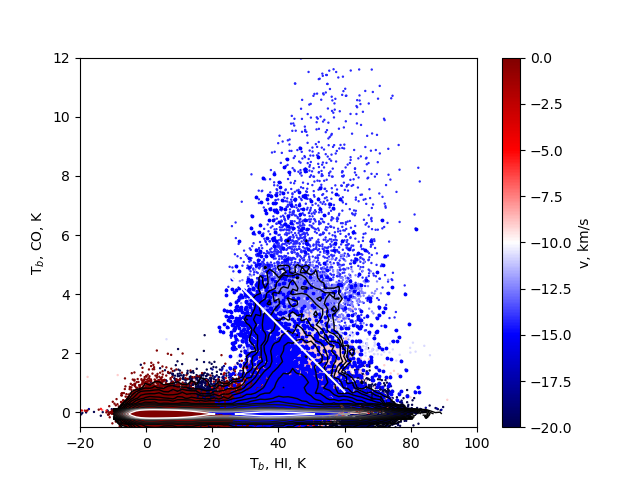}
            \caption{Scatter plot of the \Hi\ vs the $^{12}$CO emission. The red-blue colour indicates the {channel} velocity. The scatter plot is overlayed with A 2D histogram {shown by black contours}. The possible correlation {within the} [--15.1, --16.9]~\kms\ {range} is {shown} AS A white line.}
            	\label{fig:hi_to_co}

    \end{minipage}
    \hfill
        \begin{minipage}[t]{0.48\textwidth}
        	\includegraphics[width=\linewidth]{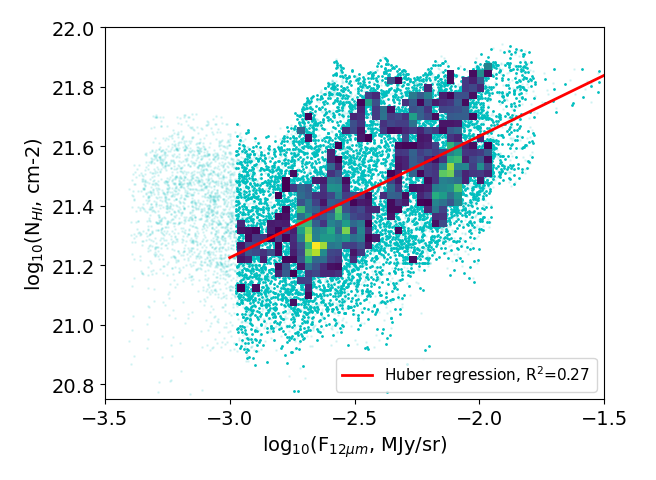}
        	\caption{The correlation between {the} 12$\mu$m WISE flux and \Hi\ column density.}
        	\label{fig:colden_12}
    \end{minipage}
\end{figure*}

{By this view} S187 {region} is located at the intersection of the bubbles. {t}he molecular gas appears only {at $\sim-$12.5 \kms,} on one side of the bubbles.

\subsection{The shell}\label{sct:shell}

Using the data discussed {in Sect. \ref{sct:results}} we traced different parts of the expanding gas and dust surrounding the \Hii\ region. The overall picture is shown in Fig.~\ref{fig:features}. The channel maps of the \Hi\ emission compared with the CO isotopologue lines are shown in Figs.~\ref{fig:hi_12co_chmap}, \ref{fig:hi_c18o_chmap}, \ref{fig:hi_13co_chmap}. Our observations indicate an expansion of the atomic {gas} {related to} the shell. The expanding part has {an almost circular} shape. It is centered {on} the \Hii\ region. The position-velocity diagram {is shown} in Fig.~\ref{fig:shell_pv}{. The emission peaks can be associated with the shell walls}, which are brighter near  molecular emission peaks. The velocity of the H{$109$}$\alpha$ line related to the \Hii\ region is $-$14.6 \kms\ \citep{Joncas} {. It is close to the expanding layer centre velocity}. The \Hi\ emission related to S187 {is detected} in the $-$21.5 to $-$8.5~\kms\ range (Fig. \ref{fig:shell_pv}). The {PDR} {atomic} layer at {the centre} should expand directly towards and outwards {with respect of} the observer. {As {a} result, no \Hi\ emission should be observed {at} $\sim -$14.6~\kms\ and towards the \Hii\ region.} {However}, the \Hi\ emission is observed {towards} the {zero offset in} the PV diagram{ (see Fig.~\ref{fig:shell_pv}). This emission } cannot be explained as line broadening of the Doppler-shifted wall. The source of the emission can be the atomic gas inside or outside of the sphere. The first {possibility implies that neutral gas fragments remain within} {the} \Hii\ region{,} {that are} not involved in {a} systematic expansion. The other {possibility implies the presence of} neutral atomic gas dissociated by leaking {radiation from the \Hii\ region} that is not involved in systematic motions either.
\begin{figure}
    \centering

    	\includegraphics[width=\linewidth]{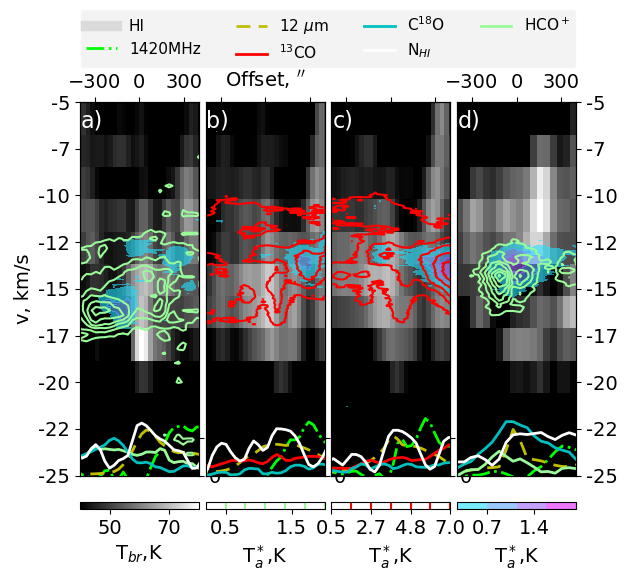}
	    	\caption{The position velocity diagrams across different {slices} of the shell {(Fig~\ref{fig:pv_maps})}.  a) - the slice through S187~PDR1 (positive offsets are towards S187B, the S187 exciting cluster). b) the slice along the direction where no wall is detected. c) PV slice {through} S187~PDR2, d) PV slice {through} S187~SE (S187-IRS).
	    	Normalised plots of the continuum emission and integrated $^{13}$CO and \ceo\ emission along the {PV} slices are plotted {at} the bottom of the panel.
	    	(Scale factors: 8~mJy{/beam}, 7~GJy~sr$^{-1}$, 17~K~\kms, 2~K~\kms, 6~K~\kms, 
	    	5$\times$10$^{21}$ cm$^{-2}$ for 1420 MHz {continuum}, 12~$\mu$m WSSA WISE, $^{13}$CO, \ceo, HCO$^+$
	        integrated intensity and \Hi\ column density, respectively). 
	    	}
    	\label{fig:pv_pdrs}

\end{figure}

The channel-by-channel scatter plots of the $^{12}$CO and \Hi\ data does not show linear correlation (see Fig.~\ref{fig:hi_to_co}), except {for} the [--17,--15] \kms\ range. In this range a sign of negative correlation is observed and shown as a white line. It {may} trace the molecular gas dissociation processes. In the {remaining} velocities, no sign of linear correlation is observed{,} probably {resulting from} the limited resolution of the molecular data. The shell’s atomic mass estimate (260$\pm$56~M$_{\odot}$) is much higher than estimated in \citet{Joncas} (70~M$_{\odot}$,under the assumption of low optical depth from the absorption profile).
Also, the absorption profiles (Fig.~\ref{fig:hi_abs}) indicate deeper absorption than reported in \citet{Joncas}{.} {Such difference arises due to the significant optical depth correction in our work. Without it both consistent.} In addition, our approach is based on the observations of very compact sources. The atomic gas is highly fragmented, so the absorbing properties may vary on small scales. To our luck, the detected fragments have the sizes much larger  (Fig. \ref{fig:clump_hist}) than the S187-1(a\&b) sizes and their separation. The estimate of the shell's atomic mass is small compared to the molecular gas mass in the shell (6400~M$_{\odot}$, this paper; 7600~M$_{\odot}$ from \citet{Arvidsson}) suggesting that most of the molecular material is intact by the dissociating processes.

The \Hi\ shell is thinner towards the dense molecular core, S187~NE. The shell correlates with the PDR traced at 12~$\mu$m (Fig. \ref{fig:colden_12}). \cite{KirsanovaDenseAtomic} note that a density-bounded PDR leads to the merging of the H$_2$ and CO dissociation {fronts}. In S187~NE the $^{13}$CO and \Hi\ peaks are separated by less than 25~arcsec~(0.2~pc) (Fig.~\ref{fig:hi_13co_chmap}), so {the shock fronts are barely separated. The HCO$^+$ and \Hi\ emission regions spatially and kinetically overlap (Fig.~\ref{fig:pv_pdrs}, a.) supposing the mixing of the fronts.} The 12~$\mu$m peak may be shifted from the \Hi\ due to a projection effect as can be seen from the position-velocity diagram in Fig.~\ref{fig:pv_pdrs},~{a}., on the bottom plots. The 1420 MHz continuum is also plotted representing the ionized region. (S187~PDR1 is traced by \Hi\ and WISE 12~$\mu$m and S187~NE is traced by HCO$^+$ and \ceo). The PV cut in the direction where no PDR and molecular wall is observed is shown in Fig.~\ref{fig:pv_pdrs},~b. The \Hi\ emission is present, but no compact features exist, the \Hi\ layer is broader than towards S187~PDR1, up to $\sim$2.4 pc. Weak \ceo\ and $^{13}$CO emissions are detected indicating a small amount of molecular material in the shell towards the location of the PV cut. The PV cut through S187~PDR2 is shown in Fig.~\ref{fig:pv_pdrs},~c. The \ceo\ emission is weaker than in the S187~PDR1, but {stronger} than in cut b). There are WISE 12~$\mu$m and 1420~MHz continuum peaks suggesting a compression of the gas. 

\subsubsection{Fragmentation} \label{sct:fragdiscuss}
{The shell has systematic motions and related kinematics, as discussed in the previous section.} In addition, we observe smaller features identified as fragments in Sect. \ref{sct:fell} and estimated their physical properties. The slope of the mass-size relation (Fig. \ref{fig:clump_correlations}) is the same as found for molecular clouds \citep[e.g.][]{fractals, fracpow}. \citet{Kritsuk} argued that supersonic turbulence fed by large-scale kinetic energy injection is at the origin of the law. The \Hi\ fragment masses are lower than masses of molecular clumps estimated by \citet{fracpow} for the same size. {This observation} may {result from} the diffusion of the atomic gas during dissociation {or transformation} to the ionized gas state. The fragments may also have significant amounts of molecular material that are not completely dissociated into atomic gas. {This sample can be incomplete on smaller scales where fragments can not be resolved. Such incompleteness will not affect the observed relations significantly since the detected fragment sizes vary {over} an order of magnitude.}
The peak of the fragment mass function at $\sim 1-3$~M$_\odot$ (Fig.~\ref{fig:clump_hist}) is very close to the peak of the prestellar core mass function in molecular clouds \citep[e.g.][]{Konyves20}. Most probably these fragments in the \Hi\ shell represent remnants of the molecular clumps where molecular hydrogen was dissociated by the UV radiation of the central {star cluster}.

The velocity dispersion in the fragments increases with the fragment size (Fig.~\ref{fig:clump_correlations}) resembling one of the well-known ``Larson's laws'' \citep{Larson81}. However the slope of {our} relation is steeper and the absolute values of the velocity dispersion are much higher than found by Larson for molecular clouds and condensations of the same size. They are also significantly higher than in massive protostellar cores \citep[e.g.][]{Zin00-corr}. {The mass-size relation for the optically thick limit are closer to those presented in \citet{fracpow}, especially for larger fragments.} Such differences {are} in agreement with these fragments being the eroded material dissociating into the surroundings with higher turbulent or thermal motions than the ordinary molecular clumps. {The erosion may be caused by photoevaporation \citep{photoevo}.}
The high velocity dispersion leads to high values of the virial parameter for most fragments $\alpha\ga 2$ (Figs.~\ref{fig:clump_hist}, \ref{fig:clump_correlations}), which implies that they are gravitationally unbound \citep[e.g.][]{Kauffmann}. There is a trend of increasing virial parameter with fragment size. The most compact fragments are relatively dense and may represent gravitationally bound objects. 

{Alternatively, such fragments may form in PDR itself by dynamical compression. The shock can be caused by the external high pressure from regions heated by far ultra violet radiation. The fragments then dissociate and/or evaporate and loose mass. The timescales should be close to the estimated kinematical age of the sub-bubble \citep{Gorti}. During the expansion of {the} \Hii\ zone, the new material should be involved in the process so the variety of the physical parameters, especially densities and virial parameters should be observed. The S187 region is fairly young ($\sim5\times10^5$yr), but no radial dependence of the fragment properties is observed. It may be caused by the observational limits or projection effect. Still, the dependence of density and virial parameters on the fragment size demonstrate a larger scatter than the mass dependence on size (Fig.~\ref{fig:clump_correlations}). }

\subsection{The stages of star formation}
\label{sct:details}
\subsubsection{The sub-bubble}
\begin{figure}
    \centering

    	\includegraphics[width=\linewidth]{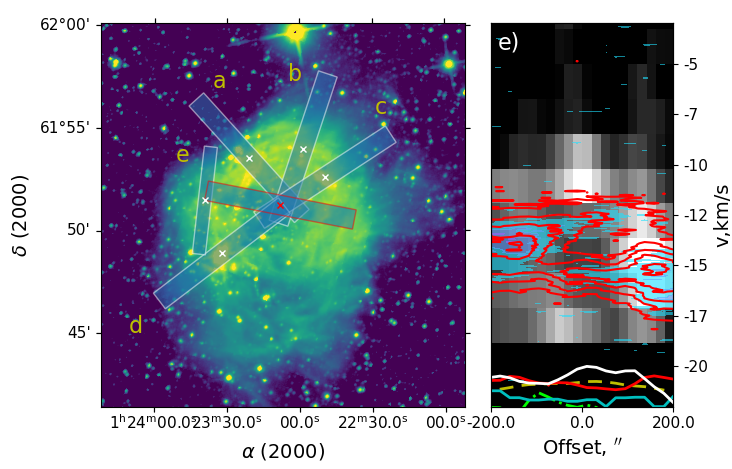}
	    	\caption{
	    	(Left:) {Locations} of the  PV slices {used in} Fig.~\ref{fig:pv_pdrs} (white boxes, with corresponding labeling) and in Fig.~\ref{fig:shell_pv} (red box). The WISE 3.4~$\mu$m image is in the background. (Right:) PV slice through cut e, the \Hi\ bubble. The {scales{, legend} and colour bars} are the same as in Fig. \ref{fig:pv_pdrs}.
	    	}
    	\label{fig:pv_maps}

\end{figure}

\begin{figure}
	\includegraphics[width=\linewidth]{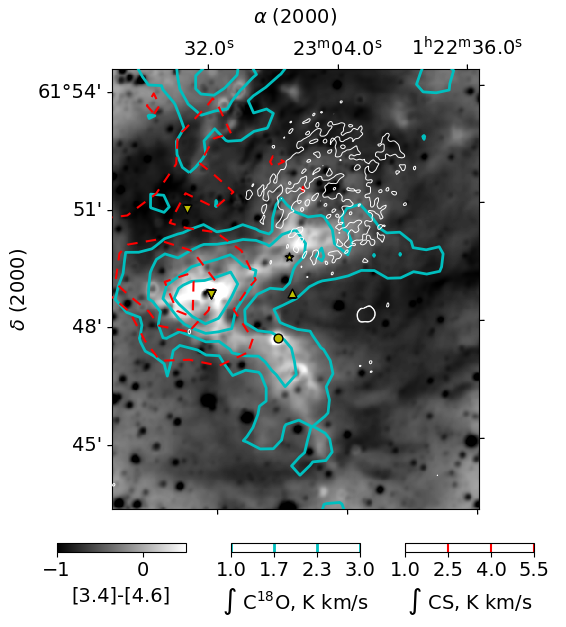}
	\caption{The molecular gas structure in \ceo\ integrated emission (cyan contours), CS (2--1) integrated emission in red dashed contours and the tapered 1420~MHz continuum in white (levels: 1, 2 mJy/beam) contours over the {background gray-scale image representing }{[3.4]--[4.6] colour}. {Point sources are not removed for presentation purposes.}
	The marks are the same as in Fig.~\ref{fig:features}}
	\label{fig:ir_43}
\end{figure}

One {of the} most complex spatial-kinematical feature in the S187 shell is the bubble-like \Hi\ structure shown in Fig.~\ref{fig:features} (labelled  \Hi\ bubble, the coordinates are given in Table. \ref{tab:molecular}).{\Hi\ channel maps and PV diagramm reveal the presence of the bubble-like structure that is not detected in other tracers.} The spectra are shown in Fig. \ref{fig:specs}. Weak HCO$^+$ and CS lines are detected. The HCO$^+$ line profile {resembles} the $^{12}$CO line profile towards {the} sub-bubble. The \ceo\ and CS peaks are located at --13~\kms\ The position-velocity diagram is shown in Fig.~\ref{fig:pv_maps}, e). The centre of the position-velocity diagram corresponds to the Class II source position, and the slice direction is shown in Fig.~\ref{fig:pv_maps}, left. The 21-cm line emission shows the O-like PV feature tracing the expanding \Hi\ layer around the Class II object with the velocity range from --19~\kms\ to --8 \kms. The peak of the 12~$\mu$m emission is close to the centre of the `O' ``hole'' in the \Hi\ data. The PV structure of the molecular lines is similar {to the \Hi}, but the velocity dispersion is lower {than for} the 21-cm line tracing outer parts of the sub-bubble. No continuum emission is detected towards the Class II object or the sub-bubble at 1.4~GHz in our data or at 8.2 and 4.9~GHz \citep{snell}. The OH maser is reported in \cite{Engels} near the Class II source. Two molecular line components exist in the $^{13}$CO, \ceo\ and HCO$^+$ emission at --15 and --12.5~\kms\ with the approximate same structure as the atomic gas. The CS integrated emission has a weak bowed structure in the direction of the boundary of the sub-bubble, as shown in Fig.~\ref{fig:ir_43}. The \Hi\ layer has higher intensity towards the \Hii\ region. The \ceo\ and $^{13}$CO position-velocity structure correlates well with the O-shape. The bubble-like structure is shown in Fig. \ref{fig:pv_maps}. As shown in \cite{Joncas} the stellar wind from the exciting cluster is pointing towards the bubble which may have triggered the early, relative to the S187~NE and SE cores, evolution. Based on the formula from \cite{Weaver} we estimate the age of the bubble as 6$\times$10$^4$~years.  

This sub-bubble structure reveals an interaction of the Class II object {with} ambient gas {within} the context of the  \Hi\ shell. The location of the Class II object near the border of the ionized region suggests A triggered origin {for} the exciting star. A number of YSOs are located towards the sub-bubble, especially at the boundary of S187. Some of them have X-ray emission counterparts reported in the Chandra point source 2XCO catalog \citep{Evans}. The excess of 12~$\mu$m emission and the 3.4 to 4.6~$\mu$m {color} detected towards the {sub}-bubble (Figs. \ref{fig:continuum_spitzer} and \ref{fig:ir_43}) imply additional production of PAHs by the exciting star and the dominance of photo dissociation near the Class-II source. The Class II source and the associated sub-bubble can lead to the additional stimulation of the YSO formation in the S187 shell environment.

\subsubsection{The S187~SE core}
The molecular core S187~SE is located to the south-east of the ionizing cluster S187B (Table~\ref{tab:molecular}). S187~SE is associated with the bright IR Class-I source \iras\ and maser sources. The position-velocity diagram across the core is plotted in Fig.~\ref{fig:pv_pdrs},~d. 
{We define the cores as kinematically-coherent gas detected in \ceo. The internal structure of this object is quite rich. The HCO$^+$, CS, HCN and N$_2$H$^+$ lines were observed by \cite{Zinchenko2009} and revealed complicated gas structure. The extended structures {are} similar to {those} observed in \ceo. Therefore, the core contains at least two dense gas fragments (probably associated with \iras\ and S187 H$\alpha$). Ammonia emission is also detected towards S187 H$\alpha$ \citep{Zavagno}.  There is a peak in the SCUBA 850 $\mu$m image located towards S187~NIRS1 {(See Fig.~\ref{fig:continuum_spitzer})} suggesting the embeddance of this object into the S187~SE core. No associated peak in molecular lines was detected towards S187~NIRS1.
}

We detected compact continuum emission which possibly traces an UC~\Hii\ region around \iras. The \Hi\ line {is} less intense towards the centre position at --15~\kms. The HCO$^+$ and \ceo\ lines have different profiles (Fig. \ref{fig:specs}) and spatial distributions (Fig. \ref{fig:pv_pdrs},~{{d}}): the \ceo\ emission peak is located towards \iras. The HCO$^+$ peak at --14.3\kms\ is shifted by $\sim$80~arcsec from the \ceo\ (--13.6 \kms) peak. {The 4.6/3.4~$\mu$m ratio peak} is located towards the HCO$^+$ peak, the CS emission peak is also shifted from \iras\  {(Fig.~\ref{fig:ir_43})}. We interpret such differentiation as an interaction between the UC \Hii\ region and the parent molecular core. The column density traced by \ceo\ associated with the core is centered at \iras\ where the dense part of the core is perturbed by the UC \Hii\ region. No self-absorption feature in HCO$^+$(1--0) (E$_u\approx$4.28K) is detected. The self-absorption was reported in HCO$^+$(3--2) (E$_u\approx$17K) \citep{Kang2012}. In case {the gas is thermalized, this would be in} agreement with a kinetic temperature of 9.5 K for the core.

No $^{12}$CO emission peak is detected towards S187~SE. The emission can be optically thick, unlike S187~NE. The emitting region towards S187~NE is more extended than the \ceo. The $^{12}$CO-traced regions, probably, represent the lower density molecular gas around the core. For S187~SE {a different environment} is observed: the \ceo,  $^{13}$CO and HCO$^+$ emission regions are surrounded by \Hi\ emission {(see Fig.~\ref{fig:pv_pdrs},~d)}. The core is embedded into the shell. Thus, the \Hi\ gas should be produced during photo-dissociation. S187~SE is dense (1$\times$10$^{5}$ cm$^{-3}$ \cite{Zinchenko2009}), and bright 12~$\mu$m spots (like S187~PDR1 and S187~PDR2) should appear at the border between PDR and molecular gas, but none was detected. The molecular material traced by \ceo\ is elongated to the centre of the \Hii\ region with the peak near the border of the \Hi\ layer. The orientation of the elongation seems close to the {plane of the sky} since no significant gradients are detected in the molecular data position-velocity diagram (Fig.~\ref{fig:pv_pdrs},~d). The distribution of the radio continuum emission indicates pocketing of the molecular material in the \Hii\ region too. The increasing ratio of the 160~to~12~$\mu$m emission towards the CS and \ceo\ emission region can be a tracer of the cold dust dominance over the core.

The origin of {the} {[3.4]-[4.6] VS \ceo\ correlation (Fig.~\ref{fig:ircorr} and \ref{fig:ir_43})} in S187~SE remains unknown. The presence of recombination lines \citep{Peeters} can increase the 4.6{$\mu$m flux} but can {only} be related to the UC \Hii\ zone. The H$_2$ lines {present in the to 4.6 $\mu$m {WISE} band also} can {increase} the colour. H$_2$ emission was observed by \citet{Salas} but shows {a} different morphology than the {colour} map. The 3.4~$\mu$m emission can be saturated by a significant optical depth in the dense core at this wavelength \citep{Pavlyuchenkov} both in PAH, silicates and grains. The optical depth at 4.6~$\mu$m is lower so such reddening can arise. {The [3.4]-[4.6] colour values is close to those presented in \citet{Ybarra} ([3.6]-[4.5] Spitzer) colour implying the reddening of the PDR emission in the S187 SE core.}

The kinematical-morphological structure of the core and a number of active star formation processes {are} indicators that S187~SE has suffered from external pressure. The most obvious (and realistic) explanation is the pressure caused by \Hii\ region expansion. Still, no shock or ionisation fronts in PDRs that should be traced in 12~$\mu$m or 3.4~to~4.6 $\mu$m ratio are detected. The core is elongated towards the center of \Hii\ zone, and the pressure {and ionizing flux} can be reduced by the geometry of the core, leading to perturbation instead of dissociation. Strong dust absorption can also protect molecular material from dissociation under ionizing cluster radiation.

\subsubsection{The S187~NE core}

The other core (S187~NE) has a mass close to S187~SE (coordinates in Table. \ref{tab:molecular}). No PDR traced by 3.6 or 4.6$\mu$m emission peaks observed under the \ceo\ contours  {suggesting} that the core is mostly intact from the expansion of the \Hii\ region. The PDR is located between the \Hii\ zone and the molecular gas, as shown in Fig.~\ref{fig:features} (S187 PDR1).  {The structure of t}he PDR, \Hi\ and molecular layer{s} {are} shown in {a} PV (Fig.~\ref{fig:pv_pdrs},~a) {plot}. The transition between the unperturbed and the affected by \Hii\ zone material is seen, with the HCO$^+$ line broadening overlapping with the \Hi\ emission. Class-I and Class-II sources are located between the S187~{NE} core and S187 \Hii\ region, but no IR sources are detected in S187~NE according to the WISE images. 

The molecular spectra of S187~NE are shown in Fig. \ref{fig:specs}. Unlike the other cores and PDRs, the HCO$^+$ and $^{12}$CO line profiles are close to the  \ceo\ ones, at --16.4~\kms. No CS peak is observed within A 1~K sensitivity. The CS(2--1) and HCO$^+$(1--0) transitions {upper energies are} close {(}7.1 and 4.28~K, respectively \citep{Shirley}{)}. The critical density of the CS(2--1) transition (1.3$\times$10$^5$~cm$^{-3}$) is two times higher than {for} HCO$^+$(1--0) (6.8$\times$10$^4$~cm$^{-3}$) \citep{Shirley}. The absence of CS may indicate A low density for this core or chemical processes varying the abundance.

{The }PV cut in Fig. \ref{fig:pv_pdrs},~a indicates an increase of the FWHM {of the} \ceo\ line from 1.4  to 3.6 \kms\ towards the ionized region. This effect can trace the turbulent motions caused by the expansion of the S187 shell. This effect {by similar effects as for \Hi\ fragments (See Sect.\ref{sct:fragdiscuss})}. The profile is far from {G}aussian, as shown in Fig. \ref{fig:specs}. The HCO$^+$ line emission (Fig.~\ref{fig:pv_pdrs},~{a}) has two components. One is a compact narrow peak shifted by $\sim$60~arcsec from the \ceo\ peak outwards of S187~B. The wider ($\sim$5\kms) weak emission correlates with the \Hi\ emission and is shifted towards S187~B by $\sim$80 arcsec. We associate this component with the photo-dissociation processes, consistent with modelling \citep{KirsanovaYetAnotherHIIModelling}.  

The $^{12}$CO peak is shifted by 40~arcsec from the \ceo\ peak to the centre of the \Hii\ region (Fig.~\ref{fig:hi_12co_chmap}). Since the $^{12}$CO emission can trace the kinetic temperature \citep{willson} this effect may be explained by heating of the core by the PDR. The PDR is brighter towards the core indicating feedback from the core to the shell. Another possibility is {an} optical depth effect, which should be greater towards the center of the core. The free-free continuum is also intense in this direction simultaneously with the peaks in the \Hi\ line. The blue spot in Fig.~\ref{fig:ir_colors} is located just between the cold dust layer (green) and the PDR (red). This spot is well-correlated with the atomic gas emission in the 21-cm line and the HCO$^+$ extended feature. This excess could be related to the shocked gas excited in Br$\alpha$ 4.06~$\mu$m line, as observed in Orion Bar \citep{Verstraete}. 
This region was observed in the H$_2$ line \citep{Salas} and no associated peak was reported. We also note that no extended green object is reported in S187~NE.

{Some} \ceo\ emission peaks {were} also detected {in} the northern and western part of the shell, as seen in Fig.~\ref{fig:hi_c18o_chmap}. The coordinates are $\alpha$=1$^h$22$^m$59$^s$.3, $\delta$=+61$^\circ$42\arcmin28{\farcs}8 and 1$^h$22$^m$41$^s$.1, +61$^\circ$49\arcmin54{\farcs}3. They are not covered by CS and HCO$^+$ observations and are not discussed in this paper.

\subsection{The PDRs}

Our observations trace the structures of dissociation regions under different physical conditions. The structures are represented in the position-velocity diagrams (Fig.~\ref{fig:pv_pdrs}) and in maps (Fig~\ref{fig:features}, left). The general structure is close to that presented in \citet{Tielens, Gorti}. 
The close-up map of the S187~PDR1 and S187~PDR2 is shown in Fig~\ref{fig:features}, {\textit{b)} and \textit{c)}}. 

We observe free-free continuum, 12$\mu$m, \Hi\, CO, HCO$^+$ and \ceo\ emission peaks sequentially from the S187B exciting cluster to the S187~NE {core} in S187~PDR1. The spatial-kinematic structure suggests {an} edge-on configuration of the PDR {(Fig.~\ref{fig:features}, \textit{c)}, Fig.~\ref{fig:pv_pdrs}, \textit{a)})}. The HCO$^+$ line is observed in {the} molecular core and towards the wall traced {by} the \Hi\ line, as shown in Fig~\ref{fig:pv_pdrs}, a, in the center of the picture. The line width increases towards the S187~PDR1.  The line profile is far from Gaussian and has multiple peaks (Fig.~\ref{fig:specs}). No absorption profile is observed towards the molecular core S187~NE (unlike for S187~SE, \citep{Kang2012}) implying the same towards S187~PDR1. No \ceo\ or SCUBA 850$\mu$m emission is detected towards the HCO$^+$ peak in S187~PDR1. The line centroid velocity is $\sim$15~km/s. 

The HCO$^+$ emitting gas is probably located behind the shocks (5.8 $\mu$m Spitzer) and the atomic gas shell layer (See Fig.~\ref{fig:features}, c.). This picture does not match the classical \citep{Sternberg} view of the evolution of PDR region, where atomic gas is located between \Hii\ and the molecular material. The structures observed at 5.8 $\mu$m look filamentary, unlike in S187~PDR2. {A} Class-II source is detected towards the S187~PDR1 HCO$^+$ line peak. The gas associated with that peak may survive after the shock passage. The HCO$^+$ abundance should increase under this process. But, there is no direct evidence of such scheme and A high-resolution multi-line analysis is necessary to study the state of the molecular gas in S187-PDR1. The difference between HCO$+$ and H$_2$ dissociation fronts is modeled in \citet{KirsanovaDenseAtomic}. Such difference may be increased by fragmentation of the gas.  The PV diagram of the \Hi\ line is not symmetric relative TO the molecular line peaks. Molecular lines are brighter at the positive velocities. The \Hi\ line peak is slightly shifted to negative velocity. Since the atomic gas is generated during dissociation, the effect may be caused by non-uniform irradiation or weakening by the HCO$^+$ traced gas fragment.

According to PV diagrams {(Fig.~\ref{fig:pv_pdrs}, \textit{c)})}, the S187~PDR2 region is more or less edge-on {with respect} to the observer ($\sim 20^\circ$). A weak \ceo\ emission region associated with SCUBA 850~$\mu$m peak is located at -12~km/s 
(Fig.~\ref{fig:pv_pdrs}, c, Fig.~\ref{fig:features}, b. and Fig.~\ref{fig:continuum_spitzer}) near S187~PDR2. Structure of the 5.8~$\mu$m Spitzer and the 4.6~$\mu$m WISE emission suggest shielding by molecular material. The column density map (Fig. \ref{fig:hi_colden}) shows that the atomic gas is concentrated near the \Hii\ zone border. 
The \Hi\ PV diagram indicate presence of the border towards the \Hii\ zone, same as for (Fig~\ref{fig:pv_pdrs}, b) where no shielding by molecular material is observed. The PV diagram (Fig.~\ref{fig:pv_pdrs}, {b}) and channel maps (Fig.~\ref{fig:hi_12co_chmap},\ref{fig:hi_c18o_chmap},\ref{fig:hi_13co_chmap}) trace leaking \Hii\ zone emission trailing up to 4 pc from the center of {the} \Hii\ zone. {From} the spatial-kinematic structure of the \Hi\ emission we conclude that {the} \Hii\ zone is not effectively shielded by molecular gas or dust fractions.

\section{Conclusion}

We studied the {shell around the S187 \Hii\ region in the \Hi\ 21 cm line, molecular lines, and also in infrared and radio continua}. The region is mostly circular, but the emission is not {homogeneous}. The {1420 MHz} continuum emission peaks are correlated with the PDR traced by WISE and Spitzer data. We detected an excess of  \Hi\ emission towards the \Hii\ region tracing the shell. It has a shape close to circular, the peaks correlate with PDRs. The position-velocity diagram in the \Hi\ line has the O-type shape {typical} of an expanding atomic layer {in this case} at the boundary of the \Hii\ region. The shell has a molecular component visible in the $^{12}$CO (1--0) line. The \ceo\ (1--0) and \Hi\ emission distribution are spatially and kinematically related. The molecular part of the shell is more extended than the atomic one. The expansion speed is $\sim$7.5~\kms. The sandwich-like composition of the ionized-PDR-HI-molecular gas structure of the \Hii\ region and {its} environment {is {defined by} different tracers}.

\vspace{3mm}

Summary:\begin{enumerate}
\item We present the results of high-resolution 8 arcsec (0.06~pc) observations of the \Hi\ line emission towards the S187 \Hii\ region. {To our knowledge, this is the highest resolution available of the \Hi\ emission observations of Galactic regions.} 
\item We resolve{d the} \nvssrg\ {source} as two separate sources with individual \Hi\ absorption profiles. {By combining} with the emission spectra {we} estimated optical depth { profiles.} The mass of the atomic gas related to S187 is estimated as {260$\pm$56~M$_\odot$} using the emission and absorption data. The median spin temperature over the shell is $\sim$50~K. The size of the shell is $\sim$4 pc. {The \Hi\ layer has varying thickness from 0.2~pc to 4~pc.}
\item The distribution of the neutral atomic gas is close to an expanding sphere, spatially {correlating with} the WISE 12~$\mu$m emission and traces the interaction between the \Hii\ region and the molecular cloud.

\item The atomic shell is {in}homogeneous and contains $\sim$100~fragments  {of} median mass around $\sim$1.1~M$_\odot$. The  total mass {estimation is in 114-240M$_{\odot}$ (depending on optical depth assumption)} which is close to the overall shell mass. The fragment sizes vary from 0.03 to 0.23~pc and the index of the mass-size power-law fit is {2.33-2.6 range}. This estimation is close to previously reported values {available }in the literature for molecular gas clumps and clouds. {These fragments can be the dissociating molecular prestellar cores under UV erosion}{.}  {Alternatively, such fragments may be formed in the PDR by impulse heating and turbulent motions.}

\item We detect an \Hi\ sub-bubble inside the \Hi\ shell. A group of young stellar objects is detected in the direction of the walls, their formation may be induced by the interaction between the bubble and the shell. The sub-bubble is young with A {kinematical} age of $\sim$10$^5$ years.

\item {Two molecular cores (NE and SE) are present in the S187 environment. They} have similar masses {($\sim$1200~M$_\odot$ and $\sim$900~M$_\odot$, respectively)} and different physical conditions. S187~SE contains a number of young stellar objects, an IRAS source, outflows and a number of other signs of ongoing star formation, which is an argument of A past interaction with the \Hii\ zone. The spatial-kinematic relation between the \Hi\ and $^{12}$CO emission implies ongoing interaction between the atomic shell and the molecular core, but no shock signatures or PDR AT the border are observed. The S187~SE virial parameter is $\sim$2.75. S187~NE is being compressed by the shell and {its} virial parameter is $\sim$1.4. No significant star formation activity is detected inside the core. Three YSOs are present inside the \Hi\ shell contact zone. The distribution of molecular lines suggests heating and the increase of turbulence near the contact zone. The core periphery located towards {the} \Hi\ shell and {the} \Hii\ zone is eroded by {the} radiation field.

\item The WISE {[3.4]$-$[4.6] colour} correlates with the column density of the molecular gas. {One of the most} reasonable explanation is the effect of {photon} absorption  {by} grains in {the} dense molecular gas of S187~SE.
\item The \Hi\ line radial plots and position-velocity diagrams are different for the directions where the \Hii\ region is bounded by dense gas and where it is not. {The }PV diagram has an O-type shape {where the shell is density bounded}. {In} the regions where no dense structures are detected, the PV structures are irregular.

\end{enumerate}

\section*{Acknowledgements}

We are thankful to Maria Murga, Yaroslav Pavlyuchenkov and Maria Kirsanova for helpful discussions and consultations. We appreciate by Kim Ardvisson and Cris Ardvisson for providing CO isotopologues observations data. We thank the anonymous reviewer for a critical reading of the manuscript and the useful {and thougtful} comments and suggestions, which greatly improved the contents of the paper.

This work was supported by the Russian Science Foundation under project No. 17-12-01256 (the \Hi\ line, molecular observations and analysis, physical parameter estimations). 
The dust emission analysis was supported by the Foundation for the Advancement of Theoretical Physics and Mathematics “BASIS”.

The Canadian Galactic Plane Survey (CGPS) is a Canadian project with international partners. The Dominion Radio Astrophysical Observatory  is operated as a national facility by the National Research Council   of Canada. The Five College Radio Astronomy Observatory CO Survey of  the Outer Galaxy was supported by NSF grant AST 94-20159. The CGPS is supported by a grant from the Natural Sciences and Engineering Research Council of Canada.". 

We acknowledge support from {the} Onsala Space Observatory for 
the provisioning of its facilities/observational support for obtaining the HCO$^+$ data. 
The Onsala Space Observatory national research infrastructure 
is funded through Swedish Research Council grant No 2017-00648.

We thank the staff of the GMRT who have made these observations possible. 
{DKO and SKG  acknowledge the support of the Department of Atomic
Energy, Government of India, under project Identification No. RTI 4002.}
\section*{Data Availability}
The raw data underlying this article are available in the GMRT Online Archive at \url{https://naps.ncra.tifr.res.in/goa/data/search}, and can be accessed with Project Code 26\_012.

The processed data underlying this article (combined GMRT and CGPS datacube) are available in Zenodoo, at \url{doi:10.5281/zenodo.5070398}

\bibliographystyle{mnras}
\bibliography{bib.bib} 

\appendix


\section{k-Nearest Neighbor regression for \Hi\ maps fitting}
\label{appendix:fitting}
{
This appendix describes the usage of the k-Nearest Neighbors regression for the analysis of the spectral maps using the Gaussian or simple radiative transfer models.
The common way is to perform individual point-by-point least squares fitting of the model parameters. The major problem of this approach is that the results of the fitting are heavily dependent on the initial conditions of the fitting procedure. The \Hi\ spectra can consist of a number of peaks represented by individual Gaussians which lead to cases where no single  statistically important solution exists. This effect increases with decreasing signal-to-noise ratio. The model parameters and it's confidence regions can be found using the sampling of the likelihood function by Monte Carlo Markov Chains \citep{emcee}, by grid sampling of dimensionally-reduced space \citep{PCA} or other similar sampling methods. Such approach is problematic for spectral map analysis OF individual positions. The computational complexity of sampling is quite high (especially for high dimension models) which should be repeated for individual positions and likelihood functions.}

{
In the case of S187 \Hi\ data there is a position on the map where the parameters are known by analysis of the absorption spectra. Such information can be used for the map analysis. We assume that the physical parameters are spatially-correlated, and the results of the previous fittings can be used as the initial parameters for gradient descent (or similar) methods. We define the residual:}

{
\begin{equation} 
\label{eq:chi}
\chi^2=\frac{1}{N_v-N_{p}}
\sum_{i=1}^{N_v} 
{\frac {(I_{i}^{obs}-I_{i}^{mod})^2} {\sigma^2}} \hspace{2mm},
\end{equation}
}

{
where $N_p$ is the number of parameters in the model, $N_v$ is the number of spectral channels used for fitting. $I_{i}^{obs}$ is the observed intensity of the spectral channel $i$ and $I_{i}^{mod}$ is the model intensity (from the Eq.~\ref{eq:br}). We use {\sc lmfit} \citep{lmfit} procedure and  Levenberg–Marquardt algorithm to find the minimum of the residual function. In general, the initial conditions for the fitting procedure can be approximated from previous calculations in different positions of the map. We use the k-Nearest Neighbor (kNN) regression as the approximation:
\begin{equation}
    \label{eq:knn}
    p_k=\frac  {\sum_{j=1}^m p_j d(I_k, I_j)^2} {\sum_{j=1}^m d(I_k, I_j)^2} \hspace{2mm},
\end{equation}

where $p_k$ is the model parameter value, $d(I_1,I_2)$ is the distance between two spectra defined as:
\begin{equation}
    d(I_1,I_2) = \sqrt{{N_v^{-1}\sigma_1\sigma_2}\times{\sum_{j=1}^{N_v} (I_{1j} - I_{2j})^2}} \hspace{2mm},
\end{equation}
where m is the number of Nearest Neighbors used for the estimation. We use the Euclidean metric in neighbors space. This is equal to a weighted average of the model parameters from $k$ samples. These $k$ samples are chosen by minimal squared deviation from current spectra. {We normalize the distance by the number of spectral channels and the RMS deviation estimated in line-free channels. The effect of the number of channels is not important since it is the same for the map. The distance to noisier data is increased by multiplying BY $\sigma$. }We use the procedures from {\sc scikit-learn} \citep{scikit-learn} in our algorithm. {Using this} approach we reduce the initial conditions problem to the adequate sample generation. The absorption spectra decomposition results and the Least Squares fitting results was used to fill the samples. We repeat map analysis a number of times until the results converge.
}
{
The algorithm used to estimate the physical parameters of the map {follows this sequence}:
}

{
\begin{enumerate}
    \item Estimate the optical depth and spin temperature profile from the absorption and emission spectra.
    \item Fit the profile with the flat-layer model.
    \item Fit the emission spectra to the nearest positions with the initial conditions from {the} previous step{.} {Mark the positions as fitted} and put the observed spectra and the fitted results in the dataset.
    \item Fit the nearest unfitted positions with the initial conditions estimated using kNN and the dataset. {Mark the positions as fitted.}
    \item Repeat the previous step until the full map is fitted.
    \item Repeat the previous two steps {starting from positions near {the} absorption: refit every position with new initial conditions from kNN results. Repeat} until the fitted results for individual {positions} converge.
\end{enumerate}
}

{
{The aim of such iterations is to fill the training set for kNN regression. We repeat the whole map analysis until the initial conditions estimated by kNN for the least-squares fitting will be close to the fitted results. The initial conditions obtained by averaging the results from the similar spectra also reduce the computation cost. The insufficient sampling of the training set on the first iteration is negated by {the} next iterations.}
Such an approach is relatively computationally efficient because the initial conditions are close to the fitted results and the gradient descent (or similar) methods need a large number of iterations to fall into a residual minima. The only assumption is that the same physical properties produce the same spectra. Still, the interval of the confident estimation in individual fit is important and can be used for the analysis of the regions where the estimation is reasonable and where it is not.
}
\section{Figures}
\label{appendix}

\begin{figure*}
	\includegraphics[height=0.8\textwidth,angle=90]{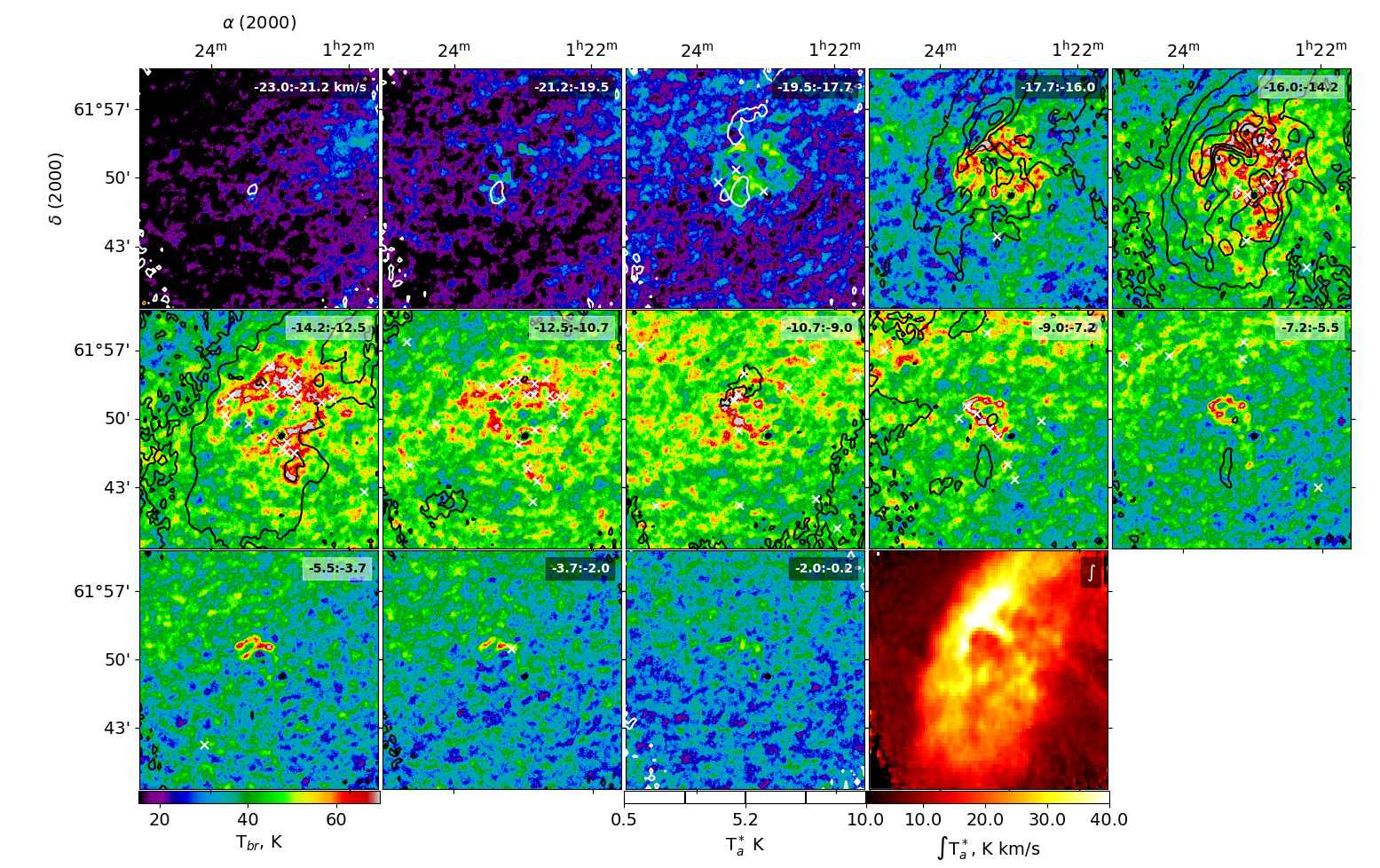}
	\caption{The channel maps of the S187 environment. The $^{12}$CO (1-0) emission is in black contours. The combined GMRT+CGPS  \Hi\ channel maps are shown in the background image. The beams are shown in lower-left corner of the first subplot. The velocity ranges are labelled at the top-right of every subplot. The integrated $^{12}$CO (1--0) {line emission}  is shown in the last subplot. The fragments identified by Fellwalker algorithm (See sect. \ref{sct:fell}) is shown as white crosses.}
	\label{fig:hi_12co_chmap}
\end{figure*}

\begin{figure*}
	\includegraphics[height=0.8\textwidth,angle=90]{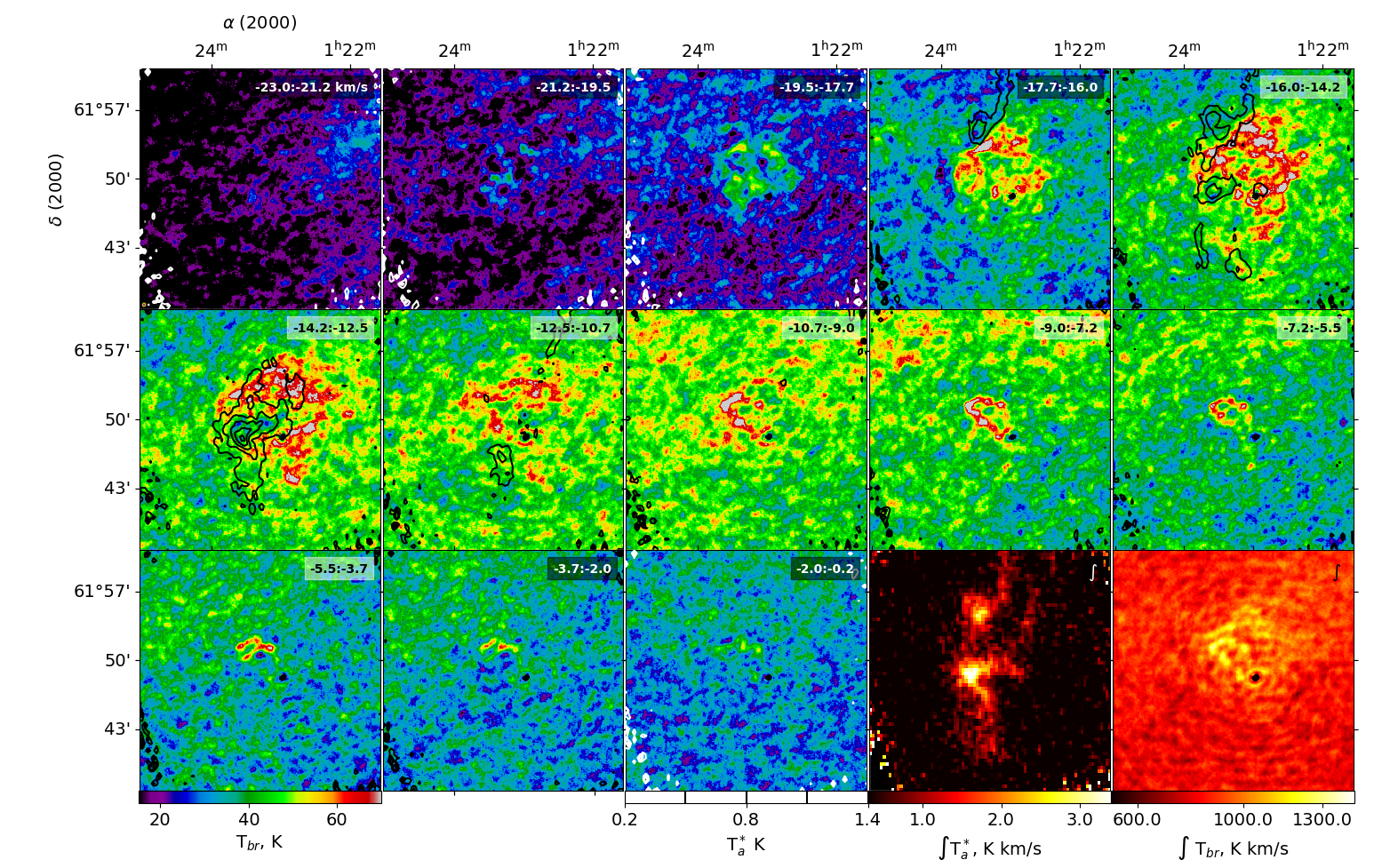}
	\caption{The channel maps of the S187 environment. The C$^{18}$O (1-0) emission is in black contours. The combined GMRT+CGPS  \Hi\ channel maps are shown in the background image. The beams are shown in lower-left corner of the first subplot. The velocity ranges are labelled at the top-right of every subplot. The penultimate last subplot represents integrated emission of C$^{18}$O (1--0) and the 21-cm line integrated emission image is shown in the last subplot. }
	\label{fig:hi_c18o_chmap}
\end{figure*}

\begin{figure*}
	\includegraphics[height=0.8\textwidth,angle=90]{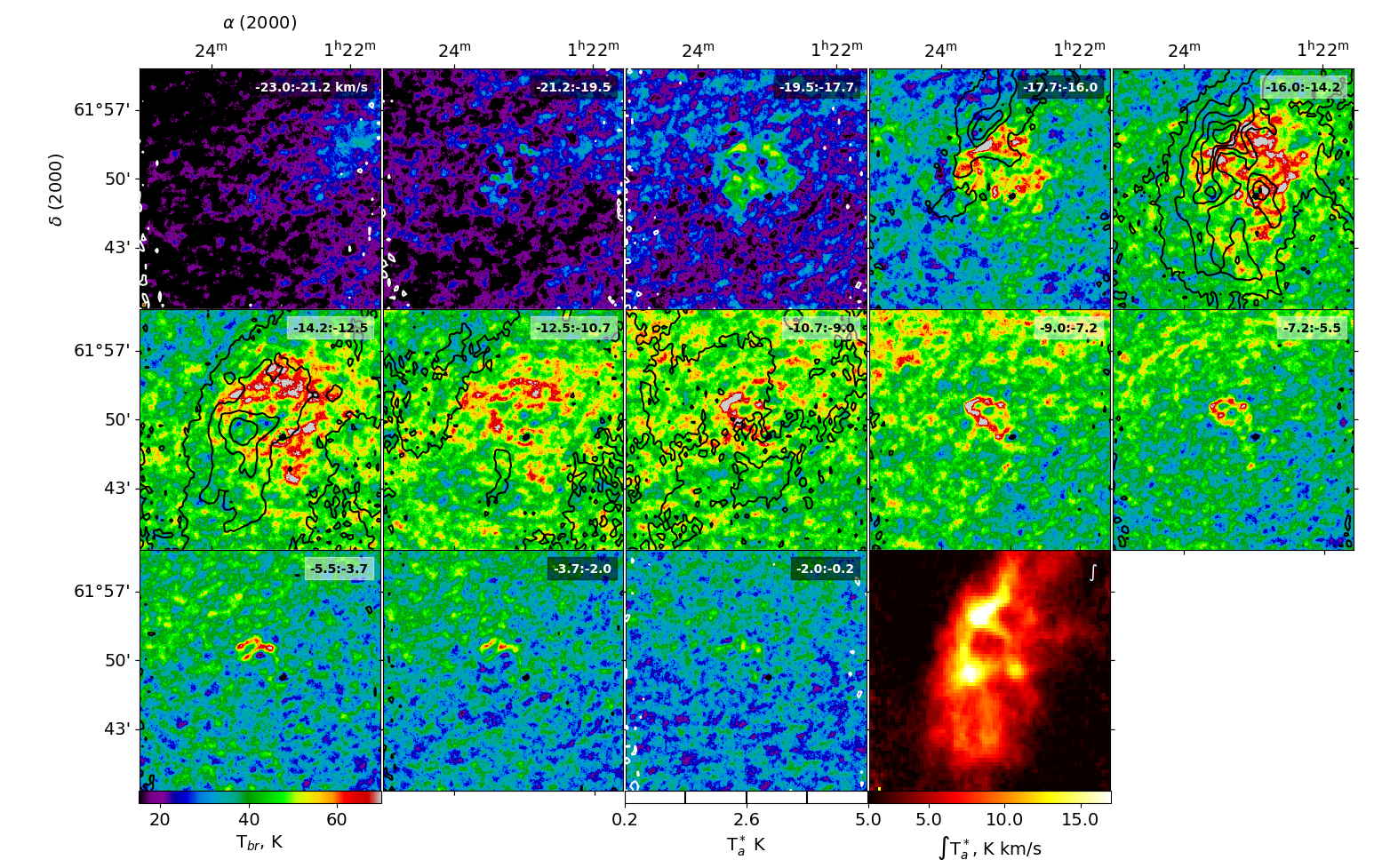}
	\caption{Same as for Fig. \ref{fig:hi_12co_chmap}, but for $^{13}$CO (1--0)}
	\label{fig:hi_13co_chmap}
\end{figure*}

\bsp	
\label{lastpage}
\end{document}